\documentclass[aps,prl,preprint,groupedaddress]{revtex4-1}
\usepackage{graphicx}
\usepackage{siunitx} 
\usepackage{textcomp}
\usepackage{soul, color}

\begin{document}

\title{Positive temperature coefficient of anisotropy due to $d^6L$ groundstate in   $\epsilon$-Fe$_2$O$_3$ }

\author{Rachel Nickel}
\email{rachel.nickel@esrf.fr}
\affiliation{Department of Physics \& Astronomy, University of Manitoba,  R3T 2N2, Winnipeg, MB, Canada}
\affiliation{European Synchrotron Radiation Facility, 38000 Grenoble, France}

\author{Kurt Kummer}
%\email{kurt.kummer@esrf.fr}
\affiliation{European Synchrotron Radiation Facility, 38000 Grenoble, France}

\author{Johan van Lierop}
\email{Johan.van.Lierop@umanitoba.ca}
\affiliation{Department of Physics \& Astronomy, University of Manitoba, Winnipeg, R3T 2N2, Canada}
\altaffiliation{Manitoba Institute for Materials, Winnipeg, MB, R3T 2N2, Canada}


\begin{abstract}

Positive anisotropy temperature coefficients ($dK/dT>0$) usually arise from hybridization between magnetic 3d states and strongly spin–orbit-coupled (SOC) subsystems. Yet $\epsilon$-Fe$_2$O$_3$ shows $dK/dT>0$ (125–200 K) without an obvious SOC partner. We study pure and Cr-doped $\epsilon$-Fe$_2$O$_3$ across the transition from low-temperature incommensurate to high-anisotropy magnetic phases. We identify a $d^6L$ ground state stabilized by itinerant hybridization in $\epsilon$-Fe$_2$O$_3$, while reduced hybridization yields $d^6 + d^6L^2$ in the Cr-doped phase, highlighting metal–ligand hybridization as a route to tune magnetic, electronic and orbitronic properties.

\end{abstract}

\maketitle

%\section{Introduction}

\emph{Introduction--} Magnetocrystalline anisotropy ($K$) is  driven by spin-orbit coupling, where an atom's electron spin is tied to its orbital motion, which is in turn shaped by the  crystal's electrostatic fields generated by the lattice. As a result,   the magnetic orientation becomes linked to the lattice, generating a preferred spin orientation in the material. In the absence of defects, domain wall motion etc.,  the magnetic coercivity of a material ($\mu_0H_C$) is proportional to $K$. In most magnetic materials, magnetocrystalline anisotropy decreases with increasing temperature ($dK/dT<0$) below their magnetic ordering temperature. However, there are a number of exceptional materials with the opposite behaviour ($dK/dT>0$), where anisotropy strengthens with increasing temperature. This unusual property makes them attractive for applications that need strong room-temperature coercivity, such as permanent magnets.

The most notable example of this is MnBi, a rare-earth element-free permanent magnet candidate, where the Mn sublattice shows the conventional $dK_{Mn}/dT<0$ while strong hybridization and spin-orbit coupling of the Bi sublattice with thermal lattice expansion results in  $dK_{Bi}/dT>0$\cite{antropov2014magnetic, antonov2020low, enkhtur2025atomic}. Because this latter term is dominant at higher temperatures,  $dK/dT>0$  for a large temperature range.

While MnBi provides the clearest example of  $dK/dT>0$ as a result of intrinsic competing anisotropies, it is far from the only material with such behaviour. Many of the rare-earth-transition-metal intermetallics (RCo$_5$, R$_2$Fe$_{14}$B where R=Nd, Sm etc.)  show $dK/dT>0$ over a finite temperature range due to competition between the $3d$ and $4f$ systems\cite{patrick2019temperature,grossinger1986temperature}, while the spin reorientation transitions of rare-earth ferrites and chromites (RFeO$_3$, RCrO$_3$) suggest a similar competition\cite{zhou2020weak, moskvin2021structure,shin2023giant}. Together these materials present a coherent picture where the $dK/dT>0$ behaviour is driven by hybridization between the magnetic $3d$ subsystem and the strongly spin-orbit coupled ``heavy'' subsystem. 

\begin{figure} [t] 
	\centering
	\includegraphics[width=0.48\textwidth]{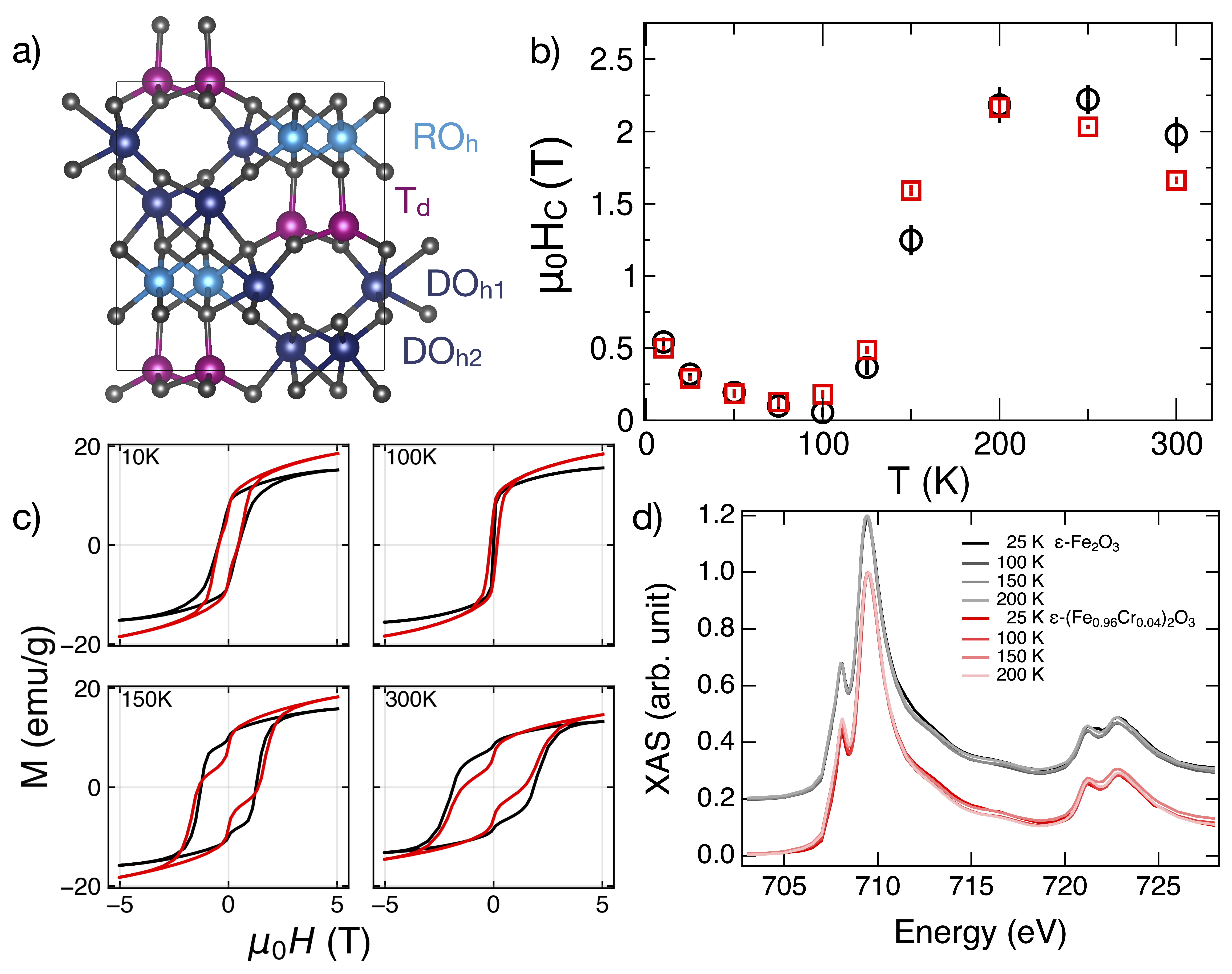}
	\caption{a) Crystal structure of $\epsilon$-Fe$_2$O$_3$. b) Temperature dependent magnetic coercivity  and c) hysteresis loops of $\epsilon$-Fe$_2$O$_3$ (black) and  $\epsilon$-(Fe$_{0.96}$Cr$_{0.04}$)$_2$O$_3$ (red) (reproduced in part from ref.~\citenum{nickel2024insights}). d) Temperature dependent  Fe L$_{3,2}$  spectra of $\epsilon$-Fe$_2$O$_3$ (black) and  $\epsilon$-(Fe$_{0.96}$Cr$_{0.04}$)$_2$O$_3$ (red), offset for clarity.  }
	\label{fig:figure1}
\end{figure} 

This model is challenged by $\epsilon$-Fe$_2$O$_3$, where $dK/dT>0$  between 125 and 250~K, but there is no clear ``heavy'' subsystem present.  Instead,  $\epsilon$-Fe$_2$O$_3$ is a perovskite with four distinct Fe sites as shown in figure~\ref{fig:figure1}a. Of these sites, one has tetrahedral coordination (T$_d$) while three have octahedral coordination (O$_h$). Two  O$_h$ sites are structurally distorted (DO$_{h}$) with varying Fe-O bond lengths, while one is not (RO$_h$). Below 100~K, $\epsilon$-Fe$_2$O$_3$ has an incommensurate magnetic state, which transforms to a hard (high anisotropy) ferrimagnetic state at the onset of $dK/dT>0$ with maximum $\mu_0$H$_C$ in excess of 2~T (see figure~\ref{fig:figure1}b,c) --similar to MnBi with its special rare-earth behaviour. This hard ferrimagnetic state  softens at high temperature (T$>$480~K) with a T$_C$ of around 850~K\cite{gich2006high, garcia2017unveiling}. 

This unique temperature dependent ordering is the result of complex electronic and magnetic interplay between the four Fe sites; however, the electronic structure of $\epsilon$-Fe$_2$O$_3$ has remained elusive. Fe L$_{3,2}$  edge x-ray absorption spectroscopy (XAS) has provided only limited insights because all cations are nominally  Fe$^{3+}$, resulting in significant overlap between site contributions\cite{tseng2009nonzero,nickel2023nanoscale}. As shown in figure~\ref{fig:figure1}d, the XAS spectra exhibit minimal differences between the incommensurate and ferrimagnetic phases.  

M\"{o}ssbauer spectroscopy has emerged as a key technique for characterizing $\epsilon$-Fe$_2$O$_3$ owing to its sensitivity to the local  Fe environment. This technique measures recoil-free absorption and emission of $\gamma$-rays to directly quantify the nuclear energy states of Fe ions. These energy levels are altered by hyperfine interactions arising from the electric charge distribution and magnetic dipole moment of  the M\"{o}ssbauer atom and from neighbouring atoms. As a result,  M\"{o}ssbauer spectroscopy provides quantitative information about the local electronic and magnetic  environment at the different Fe sites. 

Temperature dependent studies have revealed that onset of $dK/dT>0$ and the hard ferrimagnetic state occurs with the transformation of the T$_d$ local environment\cite{tronc2005spin, rehspringer2006temperature, gich2006high, kohout2015magnetic, jones2019temperature, nickel2023nanoscale}. Most notably, the magnetic hyperfine field ($B_{hf}$) of this site decreases to anomalously low values upon warming. To identify the origin of this behavior, we chose to examine both $\epsilon$-Fe$_2$O$_3$ and a perturbed system of Cr-doped $\epsilon$-Fe$_2$O$_3$ ($\epsilon$-(Fe$_{0.96}$Cr$_{0.04}$)$_2$O$_3$). Cr-doping was chosen because relative small concentrations of 
Cr dopants have been shown to disrupt the electronic and magnetic structures of other strongly-correlated perovskite systems including manganites and ferrites\cite{kimura2000variation, hong2005magnetic, raies2020temperature, kumar2021investigations, tiburcio2023influence}.

In $\epsilon$-(Fe$_{1-x}$Cr$_{x}$)$_2$O$_3$ the Cr$^{3+}$ ions dope into DO$_{h1}$ sites and act as hole defects, perturbing the electronic and magnetic structures\cite{nickel2024insights}. The $x=0.04$ doping concentration was chosen because  there is no evidence of  structural distortions or secondary phases, and the characteristic temperature dependence of $\epsilon$-Fe$_2$O$_3$ (including the temperature regime of $dK/dT>0$) is maintained.  Hysteresis loops (figure~\ref{fig:figure1}c) show a magnetically soft component increases while $M_S$ decreases with Cr doping, indicating altered Fe-Fe and Fe-O exchange interactions and hybridization. 
M\"{o}ssbauer spectra  revealed that the Cr-doping causes two T$_d$ site local environments (denoted T$_d$ and T$_d$') with hyperfine parameters that average to those of the single T$_d$ site in $\epsilon$-Fe$_2$O$_3$. Such results suggest that the Cr-doping disrupts the stabilization mechanism present in $\epsilon$-Fe$_2$O$_3$\cite{nickel2024insights}. Like the parent compound, Fe L$_{3,2}$ edge XAS of $\epsilon$-(Fe$_{0.96}$Cr$_{0.04}$)$_2$O$_3$ (figure~\ref{fig:figure1}d) shows no significant differences between the incommensurate and ferrimagnetic phases.

%\section{ Methods}
\emph{Methods--} $\epsilon$-Fe$_2$O$_3$ and $\epsilon$-(Fe$_{1-x}$Cr$_{x}$)$_2$O$_3$  nanoparticles were prepared using a metal ion-impregnated silica gel. Synthesis and characterization are described in  references \citenum{nickel2023nanoscale, nickel2024insights}. Temperature dependent L$_{3,2}$ Fe~edge and O K~edge X-ray absorption spectroscopy (XAS) was done at  beamline 4.0.2 of the Advanced Light Source. Powder samples were mounted on carbon tape, and spectra were collected in total electron yield (TEY).
Transmission M\"{o}ssbauer spectra were acquired from 10 to 300~K using a Janis SHI-850 closed-cycle refrigeration system and WissEl constant acceleration spectrometer with a $^{57}$Co\textbf{Rh} source. Source drive velocity calibration was done with a 6~$\mu$m thick $\alpha$-Fe foil at room temperature. Non-linear least squares analysis based on first-order perturbation theory was used to fit the spectra and obtain  hyperfine parameters. 
Resonant inelastic X-ray scattering was done at  beamline ID32 of the European Synchrotron Radiation Facility at 300~K. The  Fe L$_{3}$ map of $\epsilon$-Fe$_2$O$_3$ was acquired using the high throughput spectrometer at the High Field Magnet endstation\cite{kummer2016high}  with 220~meV resolution at 707~eV, while the individual spectra were acquired with the high resolution spectrometer\cite{brookes2018beamline} (30~meV resolution).  For all measurements circularly polarized light was used to maximize flux, the spectrometer was fixed at a $90^\circ$ scattering angle and acquisition was done in single-photon counting mode.

\begin{figure*} [b] 
	\centering
	\includegraphics[width=0.48\textwidth]{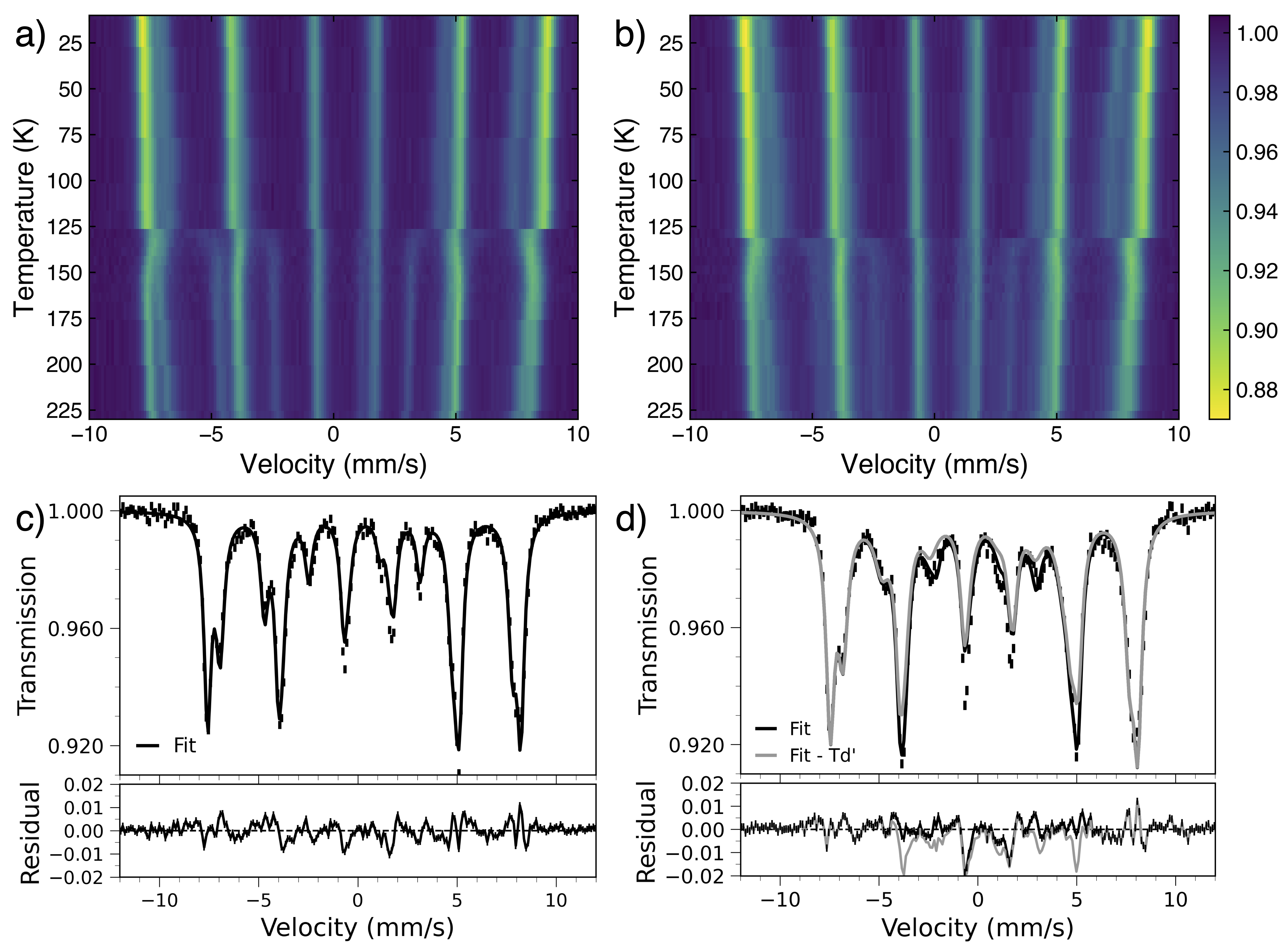}
	\caption{Maps of temperature dependent M\"{o}ssbauer spectra of a) $\epsilon$-Fe$_2$O$_3$ and b) $\epsilon$-(Fe$_{0.96}$Cr$_{0.04}$)$_2$O$_3$. Colours correspond to transmission.  200~K M\"{o}ssbauer spectrum of c) $\epsilon$-Fe$_2$O$_3$  and d) $\epsilon$-(Fe$_{0.96}$Cr$_{0.04}$)$_2$O$_3$ with fit (black line) and residuals. Grey line in d) shows the fit and corresponding residuals without the T$_d$' component.}
	\label{fig:mossy-temp-maps}
\end{figure*}

\emph{Results \& discussion --} The temperature dependent evolution of the M\"{o}ssbauer spectra of $\epsilon$-Fe$_2$O$_3$ and $\epsilon$-(Fe$_{0.96}$Cr$_{0.04}$)$_2$O$_3$ are shown in figure~\ref{fig:mossy-temp-maps}a,b. While fit parameters are essential to extract meaningful information about the individual local environments, such maps  highlight the similarities and differences between the two systems.
At low temperatures (10 to 125~K), the $\epsilon$-Fe$_2$O$_3$ and $\epsilon$-(Fe$_{0.96}$Cr$_{0.04}$)$_2$O$_3$ have remarkably similar spectra with sharp spectral lines. Between 125 and 140~K, the transition to the high anisotropy phase occurs, evidenced by discontinuities in the spectral lines as the local environments transform. Above 140~K in the high anisotropy phase, $\epsilon$-(Fe$_{0.96}$Cr$_{0.04}$)$_2$O$_3$  shows more diffuse spectral lines -- a signal of additional complexity in the local Fe environments.  Figure~\ref{fig:mossy-temp-maps}c,d show the fitted 200~K M\"{o}ssbauer spectra of $\epsilon$-Fe$_2$O$_3$ ($\chi^2 = 5.98$) and $\epsilon$-(Fe$_{0.96}$Cr$_{0.04}$)$_2$O$_3$ ($\chi^2 = 6.04$), respectively, with residuals. The black lines show the fit discussed hereafter in the text, while the grey line in figure~\ref{fig:mossy-temp-maps}d is the fit without the additional T$_d$' component, showing its importance  for fitting the broadened lines of $\epsilon$-(Fe$_{0.96}$Cr$_{0.04}$)$_2$O$_3$.

\begin{figure} [t] 
	\centering
	\includegraphics[width=0.45\textwidth]{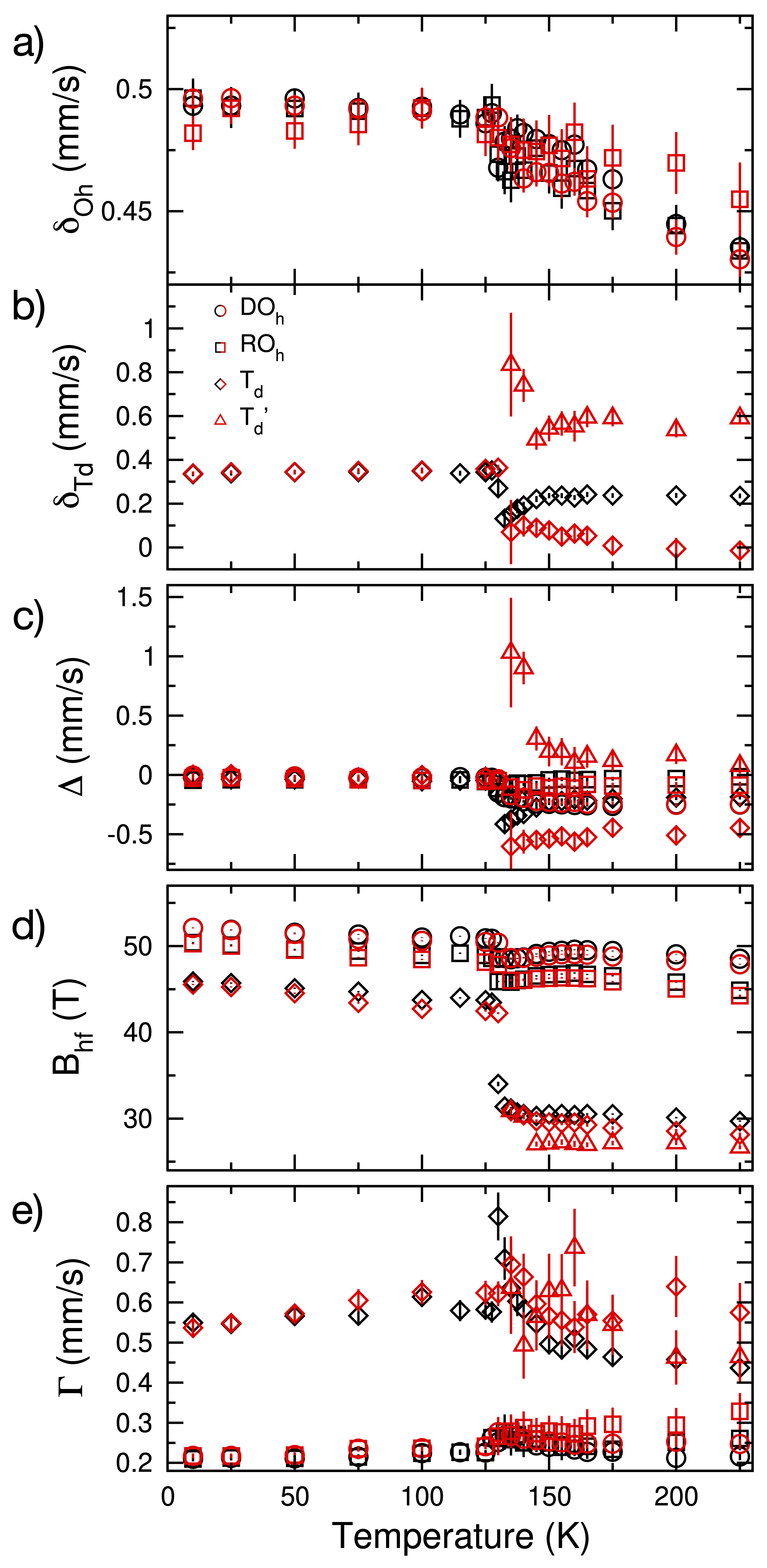}
	\caption{Temperature dependent isomer shift ($\delta$) of a) O$_h$ and b) T$_d$ components, c) quadrupole splitting ($\Delta$), d) hyperfine field ($B_{hf}$) and e) line width ($\Gamma$) of $\epsilon$-Fe$_2$O$_3$ (black) and  $\epsilon$-(Fe$_{0.96}$Cr$_{0.04}$)$_2$O$_3$ (red).}
	\label{fig:hyperfine}
\end{figure}

Keeping the overall trends in mind, we use the temperature dependent hyperfine parameters:  isomer shift ($\delta$), quadrupole splitting ($\Delta$),  magnetic hyperfine field ($B_{hf}$) and line width ($\Gamma$) shown in figure~\ref{fig:hyperfine}, to understand the evolution of the Fe local environments of  $\epsilon$-Fe$_2$O$_3$ and $\epsilon$-(Fe$_{0.96}$Cr$_{0.04}$)$_2$O$_3$.  We start in the low temperature regime ($<125$~K) to evaluate the impact of Cr-doping. The isomer shifts ($\delta$),  a measure of the $s$-electron density around the $^{57}$Fe nuclei, of all  $\epsilon$-Fe$_2$O$_3$ components (see figure~\ref{fig:hyperfine}a,b) are in agreement  with their $\epsilon$-(Fe$_{0.96}$Cr$_{0.04}$)$_2$O$_3$  counterparts. 
However, the effect of the Cr ions is evidenced by the average isomer shift of the RO$_h$ component 
decreasing, while that of the T$_d$ component 
increases. Such changes are consistent with $d$-orbital electrons shifting towards the T$_d$ site from the RO$_h$ site. Decreased quadrupole splittings ($|\Delta|$), which quantify nuclear electric field gradients,  for $\epsilon$-(Fe$_{0.96}$Cr$_{0.04}$)$_2$O$_3$ (figure~\ref{fig:hyperfine}c) reveal that the shift in electrons is a consequence of the Cr ions creating a more isotropic electronic gradient around all Fe sites. The line widths ($\Gamma$) (see figure~\ref{fig:hyperfine}e) of the corresponding spectral components of  $\epsilon$-(Fe$_{0.96}$Cr$_{0.04}$)$_2$O$_3$ and $\epsilon$-Fe$_2$O$_3$ are in agreement below 100~K. This means that the incorporation of Cr ions does not significantly increase disorder in the local environment at low temperatures. Note that the overall increase in $\Gamma$ of all components with increasing temperature is associated with increased thermal disorder. 

The reduced hyperfine field ($B_{hf}$) of the T$_d$ site (see figure~\ref{fig:hyperfine}d) provides further evidence of the more localized environment due to the Cr-doping as this  quantifies the  effective magnetic field acting on the $^{57}$Fe nuclei at a particular site and is primarily due to the isotropic Fermi contact field resulting from the spin-polarization of $s$-electrons by unpaired valence electrons. However, the anisotropic contributions to $B_{hf}$ from the dipole interactions between the valence electrons and the nuclei, and the orbital moment ($m_l$) also play a role. As the non-spherical electron density decreases, the dipolar contribution to $B_{hf}$ is reduced. While reductions in $|\Delta|$  for $\epsilon$-(Fe$_{0.96}$Cr$_{0.04}$)$_2$O$_3$ are observed in all three components, only $B_{hf,Td}$ is reduced at all temperatures. This indicates that the incorporation of Cr$^{3+}$ ions into the DO$_h$ site alters the T$_d$ site most strongly, and points to  strong DO$_{h1}$-T$_d$ site exchange.

Furthermore, while $\delta$ and $\Delta$ of $\epsilon$-(Fe$_{0.96}$Cr$_{0.04}$)$_2$O$_3$ show the same temperature dependence as $\epsilon$-Fe$_2$O$_3$ in the low temperature incommensurate phase, $B_{hf}$  diverges at 75~K. As shown in figure~\ref{fig:hyperfine}d, $B_{hf}$ of all three components in $\epsilon$-(Fe$_{0.96}$Cr$_{0.04}$)$_2$O$_3$ are reduced from those of $\epsilon$-Fe$_2$O$_3$. 
This decrease coincides with the onset of the reduction in $m_l$ that precedes the transition to the high anisotropy phase\cite{tseng2009nonzero} .The $\epsilon$-Fe$_2$O$_3$ components also exhibit a decrease in $B_{hf}$ at 100~K; however, the lower onset temperature and larger relative decrease of $B_{hf}$ highlight the reduction in $\epsilon$-(Fe$_{0.96}$Cr$_{0.04}$)$_2$O$_3$. 
This difference is caused by the reduced Fe-O hybridization from hole doping in $\epsilon$-(Fe$_{0.96}$Cr$_{0.04}$)$_2$O$_3$  because it lowers the thermal energy for the reduction of $m_l$. 
Such a shift in transition temperature is consistent with the $\mu_0H_C(T)$  shown in figure~\ref{fig:figure1}b, where  $\mu_0H_C(T)$ of $\epsilon$-(Fe$_{0.96}$Cr$_{0.04}$)$_2$O$_3$  is greater than that of $\epsilon$-Fe$_2$O$_3$  between 75 and 200~K\cite{nickel2024insights} .

Above 125~K, both $\epsilon$-Fe$_2$O$_3$ and $\epsilon$-(Fe$_{0.96}$Cr$_{0.04}$)$_2$O$_3$ reveal $dK/dT>0$   via $\mu_0H_C(T)$ as the systems transition into the high anisotropy ferrimagnetic phase. In M\"{o}ssbauer  spectroscopy this onset is observed via a  two-stage rearrangement of the local electronic structure at 127.5~K in $\epsilon$-Fe$_2$O$_3$ and 130~K in $\epsilon$-(Fe$_{0.96}$Cr$_{0.04}$)$_2$O$_3$. The most pronounced changes occur in the T$_d$ components, strongly indicating that this coordination environment drives the enhanced magnetic anisotropy. Additional insight into this high anisotropy phase are revealed in the $ \epsilon$-(Fe$_{0.96}$Cr$_{0.04}$)$_2$O$_3$ system  through  the appearance of  the second tetrahedral component (T$_d$'), indicating increased complexity within the ferrimagnetic phase.

In both $\epsilon$-Fe$_2$O$_3$ and $\epsilon$-(Fe$_{0.96}$Cr$_{0.04}$)$_2$O$_3$ $\delta_{Td}$ and $\Delta_{Td}$ (figure~\ref{fig:hyperfine}b,c, respectively) decrease above the transition. These decreases are larger in  $\epsilon$-(Fe$_{0.96}$Cr$_{0.04}$)$_2$O$_3$, but offset by the new T$_d$' component where $\delta_{Td'}$ and $\Delta_{Td'}$ are increased. In all three components the temperature dependence of $\delta$ and $\Delta$ are correlated. This is important, because while $\delta$ is commonly associated with oxidation state, it is only an indirect probe of the valence $3d$ electrons through shielding effects. More direct contributions to $\delta$  originate from changing population of the $4s$ orbitals due to bond formation as well as the compression of  $3s$ and $4s$ orbitals as a result of changing Fe-O bond lengths --the effects of which also impact $\Delta$. 

Because the temperature dependent lineshape of Fe L$_{3,2}$ edge XAS spectra does not support a change in oxidation state for either system (see figure~\ref{fig:figure1}d), the unusual parameters in the high anisotropy phase, including the   two distinct local environments of $\epsilon$-(Fe$_{0.96}$Cr$_{0.04}$)$_2$O$_3$ must be a consequence of altered Fe-O bonding. In this scenario, the smaller $\delta_{Td}$ is associated with shorter more covalent bonds, while those of the larger $\delta_{Td'}$ are longer and more ionic\cite{gutlich2010mossbauer}. Corresponding $\Delta$ behave similarly, stabilizing to -0.54(7) and 0.2(1)~mm/s for $\Delta_{Td}$ and $\Delta_{Td'}$,  respectively, with the signs revealing prolate and oblate electric field gradients. In both $\delta$ and $\Delta$, the average of the two components in $\epsilon$-(Fe$_{0.96}$Cr$_{0.04}$)$_2$O$_3$  is in agreement with those of $\epsilon$-Fe$_2$O$_3$.

The  sharp decrease of $B_{hf}$ of the tetrahedral components is the M\"{o}ssbauer signature of the transition from the incommensurate to the high anisotropy phase. Above this transition, $B_{hf,Td}$ of both $\epsilon$-Fe$_2$O$_3$ and $\epsilon$-(Fe$_{0.96}$Cr$_{0.04}$)$_2$O$_3$ become only weakly temperature dependent with average values of 30.5(1)~T and 29.7(4)~T, respectively. The T$_d$' component of $\epsilon$-(Fe$_{0.96}$Cr$_{0.04}$)$_2$O$_3$ in unique as it shows a second decrease in $B_{hf}$ at 145~K, accompanied by concurrent  reductions in $\delta_{Td'}$ and $|\Delta_{Td'}|$. The decrease in  $|\Delta_{Td'}|$ indicates a reduced electric field gradient, which weakens the dipole contribution to $B_{hf,Td'}$. Meanwhile, radial movement of the valence electrons affects both the isotropic Fermi contact term of $B_{hf}$ and $s$-electron density at the nucleus ($\delta$). This second transition, occurring 10~K above the first, acts like a correction to the initial electronic rearrangement.
Notably, this secondary transition observed in T$_d$' of $\epsilon$-(Fe$_{0.96}$Cr$_{0.04}$)$_2$O$_3$ mirrors the two-stage transition in the single T$_d$ component previously reported in $\epsilon$-Fe$_2$O$_3$\cite{nickel2023nanoscale}.  

In contrast to the tetrahedrally coordinated sites, $\delta$ of the octahedrally coordinated components of both systems  are only weakly affected by the characteristic transition and remain consistent with Fe$^{3+}$ in O$_h$ coordination. 
Instead the effect of the high anisotropy phase is reflected in changes to $\Delta$.  At the transition, $\Delta_{DOh}$ and $\Delta_{ROh}$ of $\epsilon$-Fe$_2$O$_3$ and $\epsilon$-(Fe$_{0.96}$Cr$_{0.04}$)$_2$O$_3$ become more negative (increasing in magnitude), indicating an anisotropic shift of the local electronic charge distribution. Upon further warming, the two octahedral components behave differently: $\Delta_{ROh}$ decreases in magnitude to the near-zero value expected for undistorted Fe$^{3+}$, while $\Delta_{DOh}$ becomes more negative. The near zero value of $\Delta_{ROh}$ in $\epsilon$-Fe$_2$O$_3$ indicates an isotropic electric field gradient, while $\Delta_{ROh}$ in $\epsilon$-(Fe$_{0.96}$Cr$_{0.04}$)$_2$O$_3$ is more asymmetric. Conversely, $\Delta_{DOh}$ of $\epsilon$-Fe$_2$O$_3$ is more negative than $\epsilon$-(Fe$_{0.96}$Cr$_{0.04}$)$_2$O$_3$. From these parameters we find that reducing Fe-O hybridization by doping Cr into the DO$_{h1}$ site causes the RO$_h$ (DO$_h$) site to become less (more) isotropic in the high anisotropy phase. Additionally, comparison of $\Gamma$ (figure~\ref{fig:hyperfine}e) reveals more disorder in the ferrimagnetic phase of $\epsilon$-(Fe$_{0.96}$Cr$_{0.04}$)$_2$O$_3$, particularly the RO$_h$ site. Effectively, the Cr ions prevent the $\epsilon$-(Fe$_{0.96}$Cr$_{0.04}$)$_2$O$_3$  Fe local environments from recovering --a sign of strong intersite exchange interactions in the parent compound. 

This behaviour is also identified in the temperature dependence of $B_{hf,Oh}$, which exhibit local minima at the transition temperatures of 127.5~K and 130~K for $\epsilon$-Fe$_2$O$_3$ and $\epsilon$-(Fe$_{0.96}$Cr$_{0.04}$)$_2$O$_3$, respectively, then increase with further warming to 160~K. The local minima is a consequence of spin-orbit coupling contribution being minimized as $m_l$ is minimized at the magnetic phase transition\cite{tseng2009nonzero}. The agreement between these $B_{hf}$ of $\epsilon$-Fe$_2$O$_3$ and $\epsilon$-(Fe$_{0.96}$Cr$_{0.04}$)$_2$O$_3$  is further evidence that the Fermi contact terms are not impacted by Cr-doping. The subsequent increase in the $B_{hf}$ is caused by the increase of  $m_l$ and   spin-orbit coupling as the high-coercivity magnetic phase is established and the anisotropic contributions to $B_{hf}$ increase. While $B_{hf}$ of the tetrahedral component of $\epsilon$-(Fe$_{0.96}$Cr$_{0.04}$)$_2$O$_3$ was reduced from that of $\epsilon$-Fe$_2$O$_3$ in the low temperature phase due to the incorporation of Cr$^{3+}$ ions into the DO$_{h1}$ site reducing Fe-O hybridization at the T$_d$ site, in the high anisotropy phase, all sites are impacted. This effect shows that the hard ferrimagnetic phase of $\epsilon$-Fe$_2$O$_3$ is a consequence of strong hybridization between all the Fe sites. Furthermore, since $\epsilon$-(Fe$_{0.96}$Cr$_{0.04}$)$_2$O$_3$ maintains the unique temperature dependent behaviour, it confirms that the Cr ions have a perturbative effect on $\epsilon$-Fe$_2$O$_3$ --important as we shift focus to the origin of the unusual T$_d$ environments.

Comparison of the T$_d$ component hyperfine parameters in the high anisotropy phase reveals that the average $\delta$ and $\Delta$ of  the T$_d$ and T$_d$' components in  $\epsilon$-(Fe$_{0.96}$Cr$_{0.04}$)$_2$O$_3$ map onto those of the T$_d$ component in $\epsilon$-Fe$_2$O$_3$. This suggests the T$_d$ environments observed in the Cr-doped samples are a decomposition of the single environment measured in $\epsilon$-Fe$_2$O$_3$. Since the splitting occurs with disruption to the Fe-O hybridization network, the single environment relies on ligand interactions for stability. Based on the hyperfine parameters of the three T$_d$ components, we propose an intermediate/high spin $d^6L$ groundstate (where $L$ is a ligand hole) instead of the nominal high spin $d^5$. Such a $d^6L$ character is consistent with the measured $\delta_{Td}$ of $\epsilon$-Fe$_2$O$_3$ above the characteristic transition, which is decreased from that expected for $d^5$ in T$_d$ coordination. While $\delta_{d6}>\delta_{d5}$ is true in general, the increased covalency of the $d^6L$  configuration causes the $s$ electron density to increase and $\delta$ to decrease. Anisotropy of this hybridization creates non-zero $\Delta_{DOh}$ and $\Delta_{Td}$. Most significantly, this $d^6L$ groundstate explains the characteristically low B$_{hf,Td}$ of the high anisotropy phase due to reduced Fermi contact contributions as well as the low magnetic moment (2.4~$\mu_B$) reported for the T$_d$ site\cite{gich2006high}.

\begin{figure} [t] 
	\centering
	\includegraphics[width=0.48\textwidth]{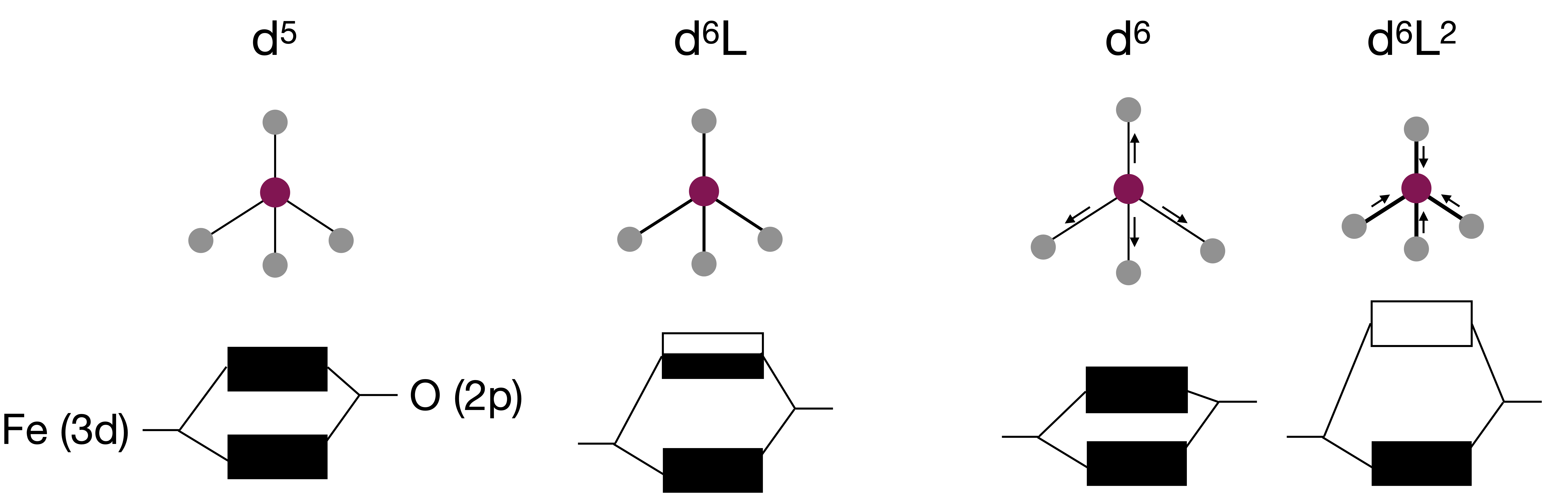}
	\caption{Schematic of electronic coupling for different electronic configurations where filled (empty) boxes  describe filled (empty) Fe-O hybrid orbitals. Distance between orbitals increases (decreases) as the coupling strength increases (decreases). Increased covalency with ligand hole formation highlighted by thicker Fe-O bonds. In the bond-disproportionated ($d^6 + d^6L^2$) picture, the more (less) ionic configuration with reduced (increased) coupling strength has longer (shorter) bonds.}
	\label{fig:schematic}
\end{figure}

A similar groundstate, with a strong ligand hole component has been linked to the perovskite SrCoO$_3$, where a $d^6L$ intermediate state is stabilized by negative charge transfer energy and a relatively large Coulomb repulsion $U$\cite{potze1995possibility,zhuang1998possible,pouchard2001spin}. Other such materials with ligand-hole dominated groundstates  include CaFeO$_3$ ($d^5L$)\cite{rogge2018electronic}, LaCoO$_3$ ($d^7L$)\cite{korotin1996intermediate} and the rare earth ($R$) perovskite nickelates $R$NiO$_3$  ($d^8L$)\cite{johnston2014charge,bisogni2016ground}. These compounds are notable because they stabilize from  $d^nL$  into a $d^n + d^nL^2$ configurations. In the case of the nickelates, the $d^8 + d^8L^2$ state is formed through bond-disproportionation\cite{mizokawa2000spin,green2016bond}. Unlike charge disproportionation, which describes a transfer of charge from one cation environment to the other, here the physical charge density around each environment is expected to remain constant while differences in bond lengths are the result of charge rearrangement around the cation\cite{dalpian2018bond}. Such behaviour is the result of a charge self-regulation mechanism, where rearrangement to the electronic density of the cation is compensated by that of the ligand\cite{raebiger2008charge}. This mechanism explains the two tetrahedral local environments in  $\epsilon$-(Fe$_{0.96}$Cr$_{0.04}$)$_2$O$_3$ where the T$_d$ and T$_d$' components describe  $d^6L^2$ and $d^6$, respectively. This $d^6 + d^6L^2$ bond disproportionated state  with  Cr-doping indicate the Cr ions disrupt the Fe-O hybridization that stabilizes the $d^6L$ groundstate in $\epsilon$-Fe$_2$O$_3$.

\begin{figure} [h] 
	\centering
	\includegraphics[width=0.98\textwidth]{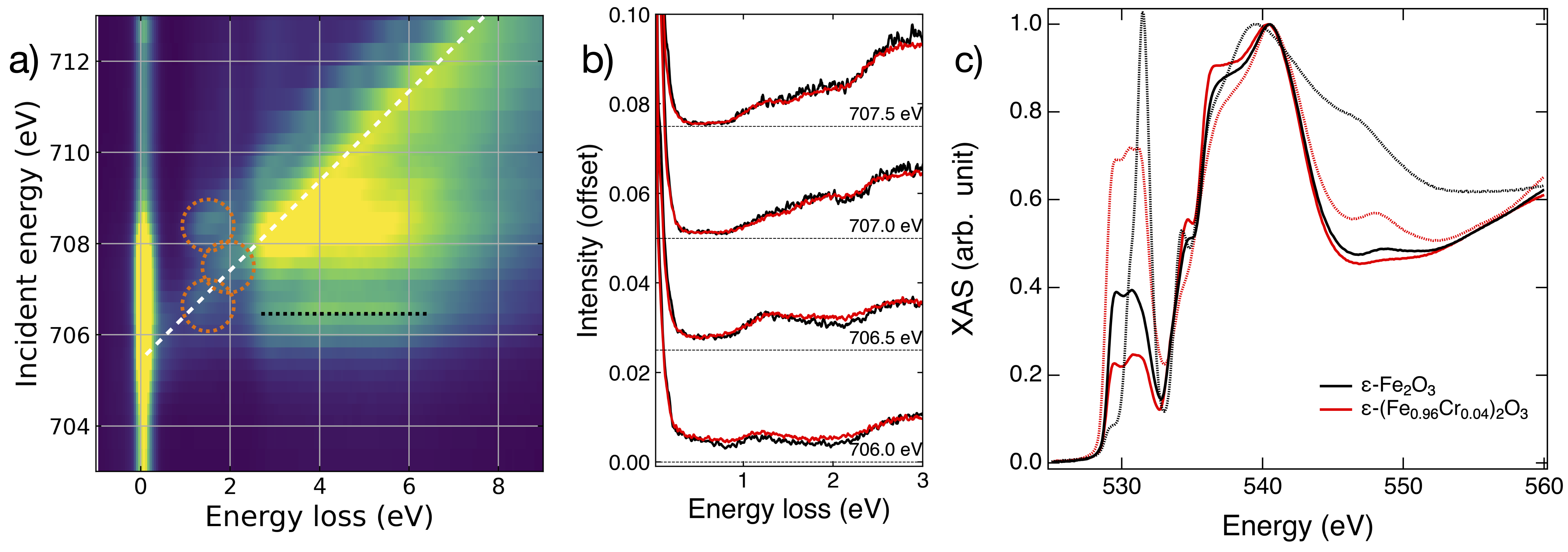}
	\caption{a)  High throughput Fe L$_3$ RIXS map of $\epsilon$-Fe$_2$O$_3$ at 300~K. Orange circles mark $dd$ excitations, while black and white lines highlight charge transfer and fluorescence, respectively. b) High resolution RIXS spectra of $\epsilon$-Fe$_2$O$_3$ (black) and $\epsilon$-(Fe$_{0.96}$Cr$_{0.04}$ (red). Spectra are offset for clarity. c) O K edge XAS  spectra of $\epsilon$-Fe$_2$O$_3$ (black) and  $\epsilon$-(Fe$_{0.96}$Cr$_{0.04}$)$_2$O$_3$ (red) at 50~K (solid) and 250~K (dashed).  }
	\label{fig:rixs}
\end{figure}

For further evidence of this negative charge transfer model, we turn to Fe $2p-3d$ resonant inelastic x-ray scattering (RIXS) at 300~K in the ferrimagnetic phase. The high throughput energy map of $\epsilon$-Fe$_2$O$_3$ shown in figure~\ref{fig:rixs}a  shows  localized $dd$ excitations, broad charge transfer excitations, and  fluorescence, as marked on the figure. Here the charge transfer excitations provide further evidence of strong O $2p$ and Fe $3d$ hybridization while the fluorescent signal extends to unusually low loss energies. 
High resolution RIXS spectra  of both $\epsilon$-Fe$_2$O$_3$ and $\epsilon$-(Fe$_{0.96}$Cr$_{0.04}$)$_2$O$_3$ (figure~\ref{fig:rixs}b) confirm that a continuum of spectral weight extends from the elastic line to the $dd$ excitations for incident energies up to 707.0~eV. Such a signal is a sign of negative charge transfer as the partially occupied O $2p$ density of states extends across the Fermi level\cite{bisogni2016ground}. In addition to the low energy fluorescence, the $dd$ excitations of the RIXS spectra are well matched between $\epsilon$-Fe$_2$O$_3$ and $\epsilon$-(Fe$_{0.96}$Cr$_{0.04}$)$_2$O$_3$. This signals the same Fe crystal field splitting, consistent with the M\"{o}ssbauer spectroscopy where variations in quadrupole splitting are on the order of 1~mm/s (48.1~neV) --well below the energy resolution of  RIXS.

While Fe L$_3$ RIXS provides evidence of negative charge transfer and highlights the similarities between $\epsilon$-Fe$_2$O$_3$ and $\epsilon$-(Fe$_{0.96}$Cr$_{0.04}$)$_2$O$_3$, O K~edge XAS (see figure~\ref{fig:rixs}c) reveals the differences. In $\epsilon$-Fe$_2$O$_3$, the spectra show a transformation in Fe-O bonding between the high and low temperature phases via a change in line shape below 530~eV (pre-peak). At 50~K the multi-peak structure is consistent with Fe-O $3d-2p$ hybridization splitting due to crystal field effects, which shifts towards a single band structure in the ferrimagnetic phase.  
More significantly, increased spectral weight at 548~eV in the 250~K spectrum indicates increased excitation into delocalized $4sp$ Fe bands\cite{leonov2006electronic}, consistent with the onset of an itinerant electronic structure connecting extended states across sublattices. This extended hybridization aids in the stabilization of the $d^6L$ configuration of the T$_d$ site and the cooperative behaviour of the different site anisotropies, leading to the the onset of  $dK/dT>0$ behaviour.

In $\epsilon$-(Fe$_{0.96}$Cr$_{0.04}$)$_2$O$_3$, the pre-peak in the 50~K spectrum shows the same lineshape as $\epsilon$-Fe$_2$O$_3$, confirming similar Fe-O hybridization in the low temperature phase, albeit weaker. 
In the ferrimagnetic phase, this hybridization is strengthened but maintains its multiband character, in contrast to the parent compound. Furthermore, the formation of the itinerant electronic structure is suppressed, as revealed by the much  smaller signal at 548~eV than  $\epsilon$-Fe$_2$O$_3$. 
The absence of an extended exchange network is correlated with the T$_d$ site splitting into  two local environments. With minimal changes to the O K~edge pre-peak line shape and the Fe L$_{3,2}$~edge XAS, this splitting is consistent with bond disproportionation rather than a transition in Fe valence or spin state. While we clearly identify a relationship between Cr-doping, bond disproportionation and a more localized hybridization, at this point we cannot determine causality.

Overall, in $\epsilon$-Fe$_2$O$_3$ the transformation of the T$_d$ site local environment signals the onset of  $dK/dT>0$ behaviour. Our multimodal results reveal the phase transition to be a spatially extended reorganization of the Fe-O hybridization  stabilizing the formation of a $d^6L$ groundstate. This reorganization affects all Fe sites, enabling the coherent addition of magnetocrystalline anisotropy. Cr-doping  the DO$_{h1}$ sites disrupts the exchange network and forms a localized bond disproportionation stabilization of the T$_d$ site. These altered  exchange interactions weaken the coupling between the different sublattices and cause an increase in the magnetically soft component. These results point to the cooperative effect of the four Fe sites in $\epsilon$-Fe$_2$O$_3$ as key to the anisotropy of the system. Being able to tune this extended hybridization would have implications for magnetic, electronic and orbitronic applications, and provides a clear direction for future studies.

\section{Appendix: M\"{o}ssbauer fitting}
The fitting model is based on first-order perturbation theory, which imposes restrictions on the components used to fit the spectra. Each component of the spectrum is fully described by five parameters: isomer shift, quadrupole splitting, hyperfine field, line width and component intensity (amplitude), while an additional parameter is used to fit the baseline of the spectrum. In total, this translates to up to 21 parameters for the fitting of our most complex spectra ($\epsilon$-(Fe$_{0.96}$Cr$_{0.04}$)$_2$O$_3$ in the ferrimagnetic phase). However, it would be misleading to call these `free' parameters as they must be restricted to physical values. 

As a nuclear decay process, the natural lineshape of M\"{o}ssbauer resonance is Lorentzian. The lower limit of the linewidth (natural linewidth, $\Gamma$) of any component is determined by the lifetime of the nuclear excited state, which in $^{57}$Fe translates to 0.13~mm/s. Increases to the linewidth reflect disorder in the local environment, but smaller values are non-physical. Importantly, all lines of a single component have the same broadening. Somewhat similarly, the relative line intensities of the sextets (which originate from Zeeman splitting) are also restricted. In the case of our powder samples, the internal magnetic field will be isotropically distributed with respect to the gamma-ray propagation. From the squares of the relevant Clebsch-Gordan coefficients, this results in a fixed 3:2:1:1:2:3 intensity ratio. 

\begin{figure} [h] 
	\centering
	\includegraphics[width=0.48\textwidth]{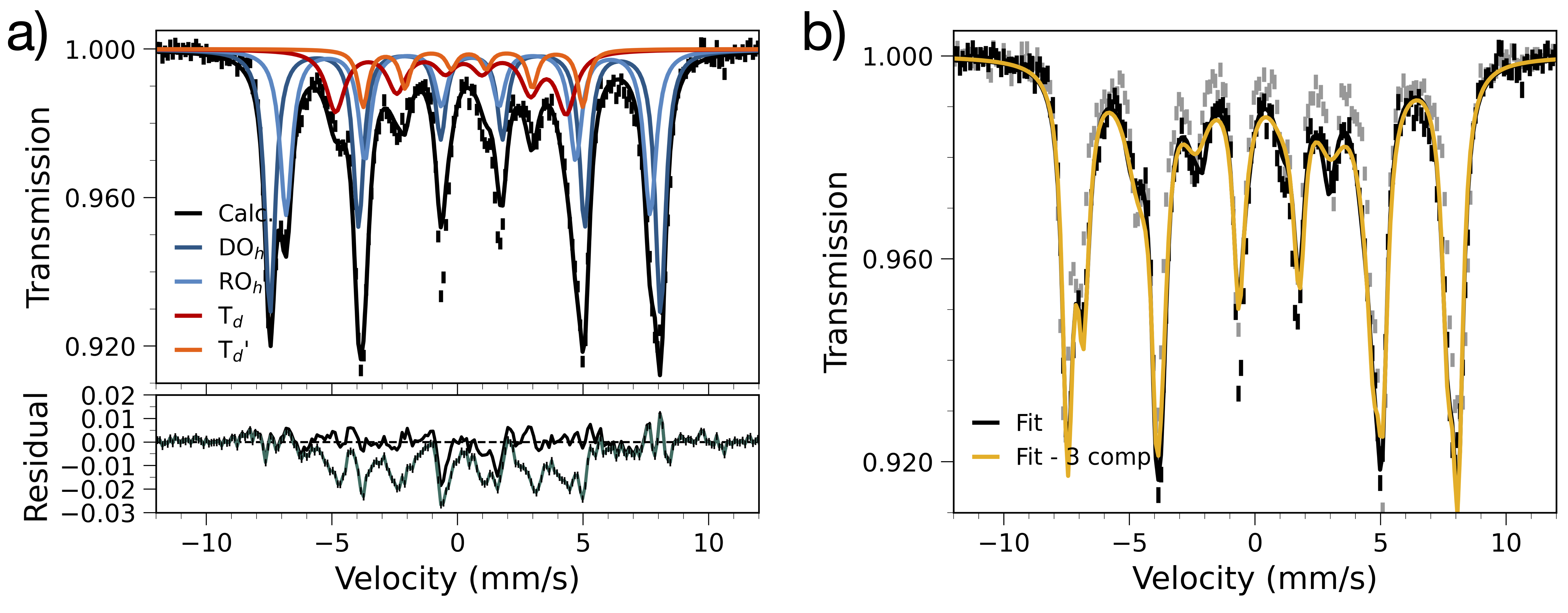}
	\caption{a) M\"{o}ssbauer spectrum of $\epsilon$-(Fe$_{0.96}$Cr$_{0.04}$)$_2$O$_3$ at 200~K overlaid with four component fit (black) and contributions of individual components. Black residual plot corresponds to total fit, while green residuals include only the DO$_h$ and RO$_h$ components as discussed in the text. b) Comparison of four component (black, $\chi^2=6.04$ ) and three component (yellow, $\chi^2=7.31$) fits. $\epsilon$-Fe$_2$O$_3$ spectrum shown in grey to highlight differences in lineshape.}
	\label{fig:appendix}
\end{figure}

The beauty of M\"{o}ssbauer spectroscopy is that many of the obtained parameters can be directly observed in the spectra. For example, the isomer shift is the centroid shift of a component (in our case a sextet) from 0, while the hyperfine field is the spacing between lines 1 and 6 multiplied by a calibration factor. The challenge comes in fitting multiple components, but here we benefit from the limits on component lineshape and relative intensities. Figure~\ref{fig:appendix} shows the 200~K spectrum of $\epsilon$-(Fe$_{0.96}$Cr$_{0.04}$)$_2$O$_3$.  To fit the spectrum, we start with the DO$_h$ and RO$_h$ sextets (dark and light blue). As these sites compose lines 1 and 6, we are unambiguously able to determine their intensities, isomer shift and hyperfine fields with minimal uncertainty, while the positions of lines 2-5 determine the respective quadrupole splittings. Thus the parameters for the first two components are fully determined. From the green residuals in figure~\ref{fig:appendix}a which include only the two octahedral components, at least one more sextet is necessary to describe the additional lines in the spectrum. However because the line at -2~mm/s is broader than that at 3.5~mm/s and has different intensity, the line shapes cannot be reproduced with a single component --in contrast to the $\epsilon$-Fe$_2$O$_3$ spectrum (shown in grey in figure~\ref{fig:appendix}b for comparison). This is a direct consequence of the line shape and relative line ratio restrictions placed by the physics of the M\"{o}ssbauer effect. The yellow fit line in figure~\ref{fig:appendix}b shows a version of the fit with only three components where the increased broadening and decreased intensity of the tetrahedral component needed to fit the region around -5~mm/s results in a poorer description of the spectral lineshape at -2 and 3.5~mm/s. Standard $\chi^2$ reflects this difference with $\chi^2=7.31$ for the three component model while the four component model gives 6.04, comparable to the fit of $\epsilon$-Fe$_2$O$_3$ ($\chi^2 = 5.98$).
Thus the four component model is favoured.

%\section{Acknowledgements}
\emph{Acknowledgements--} The authors acknowledge funding from the Natural Sciences and Engineering Research Council of Canada (Vanier Scholarship program, and RGPIN-2018-05012) and the Canada Foundation for Innovation. Use of the Advanced Light Source, a US Department of Energy (DOE) Office of Science User Facility was supported by the US DOE under contract DE-AC02-05CH11231. The authors thank the European Synchrotron Radiation Facility for providing beam time under proposals IHHC4037, IHHC4143, and  IHHC4252.

%\section{Data Availability}
\emph{Data availability--} The RIXS data are openly available at https://doi.esrf.fr/10.15151/ESRF-DC-2493094440. The other data that support the findings of this article are not publicly available. The data are available from the authors upon request.

\bibliography{Supertransferred-epsilon-bib.org.tug}

\begin{thebibliography}{37}%
\makeatletter
\providecommand \@ifxundefined [1]{%
 \@ifx{#1\undefined}
}%
\providecommand \@ifnum [1]{%
 \ifnum #1\expandafter \@firstoftwo
 \else \expandafter \@secondoftwo
 \fi
}%
\providecommand \@ifx [1]{%
 \ifx #1\expandafter \@firstoftwo
 \else \expandafter \@secondoftwo
 \fi
}%
\providecommand \natexlab [1]{#1}%
\providecommand \enquote  [1]{``#1''}%
\providecommand \bibnamefont  [1]{#1}%
\providecommand \bibfnamefont [1]{#1}%
\providecommand \citenamefont [1]{#1}%
\providecommand \href@noop [0]{\@secondoftwo}%
\providecommand \href [0]{\begingroup \@sanitize@url \@href}%
\providecommand \@href[1]{\@@startlink{#1}\@@href}%
\providecommand \@@href[1]{\endgroup#1\@@endlink}%
\providecommand \@sanitize@url [0]{\catcode `\\12\catcode `\$12\catcode
  `\&12\catcode `\#12\catcode `\^12\catcode `\_12\catcode `\%12\relax}%
\providecommand \@@startlink[1]{}%
\providecommand \@@endlink[0]{}%
\providecommand \url  [0]{\begingroup\@sanitize@url \@url }%
\providecommand \@url [1]{\endgroup\@href {#1}{\urlprefix }}%
\providecommand \urlprefix  [0]{URL }%
\providecommand \Eprint [0]{\href }%
\providecommand \doibase [0]{http://dx.doi.org/}%
\providecommand \selectlanguage [0]{\@gobble}%
\providecommand \bibinfo  [0]{\@secondoftwo}%
\providecommand \bibfield  [0]{\@secondoftwo}%
\providecommand \translation [1]{[#1]}%
\providecommand \BibitemOpen [0]{}%
\providecommand \bibitemStop [0]{}%
\providecommand \bibitemNoStop [0]{.\EOS\space}%
\providecommand \EOS [0]{\spacefactor3000\relax}%
\providecommand \BibitemShut  [1]{\csname bibitem#1\endcsname}%
\let\auto@bib@innerbib\@empty
%</preamble>
\bibitem [{\citenamefont {Antropov}\ \emph {et~al.}(2014)\citenamefont
  {Antropov}, \citenamefont {Antonov}, \citenamefont {Bekenov}, \citenamefont
  {Kutepov},\ and\ \citenamefont {Kotliar}}]{antropov2014magnetic}%
  \BibitemOpen
  \bibfield  {author} {\bibinfo {author} {\bibfnamefont {V.}~\bibnamefont
  {Antropov}}, \bibinfo {author} {\bibfnamefont {V.}~\bibnamefont {Antonov}},
  \bibinfo {author} {\bibfnamefont {L.}~\bibnamefont {Bekenov}}, \bibinfo
  {author} {\bibfnamefont {A.}~\bibnamefont {Kutepov}}, \ and\ \bibinfo
  {author} {\bibfnamefont {G.}~\bibnamefont {Kotliar}},\ }\href@noop {}
  {\bibfield  {journal} {\bibinfo  {journal} {Physical Review B}\ }\textbf
  {\bibinfo {volume} {90}},\ \bibinfo {pages} {054404} (\bibinfo {year}
  {2014})}\BibitemShut {NoStop}%
\bibitem [{\citenamefont {Antonov}\ and\ \citenamefont
  {Antropov}(2020)}]{antonov2020low}%
  \BibitemOpen
  \bibfield  {author} {\bibinfo {author} {\bibfnamefont {V.}~\bibnamefont
  {Antonov}}\ and\ \bibinfo {author} {\bibfnamefont {V.~P.}\ \bibnamefont
  {Antropov}},\ }\href@noop {} {\bibfield  {journal} {\bibinfo  {journal} {Low
  Temperature Physics}\ }\textbf {\bibinfo {volume} {46}},\ \bibinfo {pages}
  {1} (\bibinfo {year} {2020})}\BibitemShut {NoStop}%
\bibitem [{\citenamefont {Enkhtur}\ and\ \citenamefont
  {Odkhuu}(2025)}]{enkhtur2025atomic}%
  \BibitemOpen
  \bibfield  {author} {\bibinfo {author} {\bibfnamefont {U.}~\bibnamefont
  {Enkhtur}}\ and\ \bibinfo {author} {\bibfnamefont {D.}~\bibnamefont
  {Odkhuu}},\ }\href@noop {} {\bibfield  {journal} {\bibinfo  {journal}
  {Scientific Reports}\ }\textbf {\bibinfo {volume} {15}},\ \bibinfo {pages}
  {36792} (\bibinfo {year} {2025})}\BibitemShut {NoStop}%
\bibitem [{\citenamefont {Patrick}\ and\ \citenamefont
  {Staunton}(2019)}]{patrick2019temperature}%
  \BibitemOpen
  \bibfield  {author} {\bibinfo {author} {\bibfnamefont {C.~E.}\ \bibnamefont
  {Patrick}}\ and\ \bibinfo {author} {\bibfnamefont {J.~B.}\ \bibnamefont
  {Staunton}},\ }\href@noop {} {\bibfield  {journal} {\bibinfo  {journal}
  {Physical Review Materials}\ }\textbf {\bibinfo {volume} {3}},\ \bibinfo
  {pages} {101401} (\bibinfo {year} {2019})}\BibitemShut {NoStop}%
\bibitem [{\citenamefont {Gr{\"o}ssinger}\ \emph {et~al.}(1986)\citenamefont
  {Gr{\"o}ssinger}, \citenamefont {Sun}, \citenamefont {Eibler}, \citenamefont
  {Buschow},\ and\ \citenamefont {Kirchmayr}}]{grossinger1986temperature}%
  \BibitemOpen
  \bibfield  {author} {\bibinfo {author} {\bibfnamefont {R.}~\bibnamefont
  {Gr{\"o}ssinger}}, \bibinfo {author} {\bibfnamefont {X.}~\bibnamefont {Sun}},
  \bibinfo {author} {\bibfnamefont {R.}~\bibnamefont {Eibler}}, \bibinfo
  {author} {\bibfnamefont {K.}~\bibnamefont {Buschow}}, \ and\ \bibinfo
  {author} {\bibfnamefont {H.}~\bibnamefont {Kirchmayr}},\ }\href@noop {}
  {\bibfield  {journal} {\bibinfo  {journal} {Journal of magnetism and magnetic
  materials}\ }\textbf {\bibinfo {volume} {58}},\ \bibinfo {pages} {55}
  (\bibinfo {year} {1986})}\BibitemShut {NoStop}%
\bibitem [{\citenamefont {Zhou}\ \emph {et~al.}(2020)\citenamefont {Zhou},
  \citenamefont {Marshall}, \citenamefont {Li}, \citenamefont {Li},\ and\
  \citenamefont {He}}]{zhou2020weak}%
  \BibitemOpen
  \bibfield  {author} {\bibinfo {author} {\bibfnamefont {J.-S.}\ \bibnamefont
  {Zhou}}, \bibinfo {author} {\bibfnamefont {L.}~\bibnamefont {Marshall}},
  \bibinfo {author} {\bibfnamefont {Z.-Y.}\ \bibnamefont {Li}}, \bibinfo
  {author} {\bibfnamefont {X.}~\bibnamefont {Li}}, \ and\ \bibinfo {author}
  {\bibfnamefont {J.-M.}\ \bibnamefont {He}},\ }\href@noop {} {\bibfield
  {journal} {\bibinfo  {journal} {Physical Review B}\ }\textbf {\bibinfo
  {volume} {102}},\ \bibinfo {pages} {104420} (\bibinfo {year}
  {2020})}\BibitemShut {NoStop}%
\bibitem [{\citenamefont {Moskvin}(2021)}]{moskvin2021structure}%
  \BibitemOpen
  \bibfield  {author} {\bibinfo {author} {\bibfnamefont {A.}~\bibnamefont
  {Moskvin}},\ }\href@noop {} {\bibfield  {journal} {\bibinfo  {journal}
  {Magnetochemistry}\ }\textbf {\bibinfo {volume} {7}},\ \bibinfo {pages} {111}
  (\bibinfo {year} {2021})}\BibitemShut {NoStop}%
\bibitem [{\citenamefont {Shin}\ \emph {et~al.}(2023)\citenamefont {Shin},
  \citenamefont {Kim}, \citenamefont {Jeong}, \citenamefont {Kim},
  \citenamefont {Lee},\ and\ \citenamefont {Choi}}]{shin2023giant}%
  \BibitemOpen
  \bibfield  {author} {\bibinfo {author} {\bibfnamefont {H.~J.}\ \bibnamefont
  {Shin}}, \bibinfo {author} {\bibfnamefont {J.~S.}\ \bibnamefont {Kim}},
  \bibinfo {author} {\bibfnamefont {K.~W.}\ \bibnamefont {Jeong}}, \bibinfo
  {author} {\bibfnamefont {J.~H.}\ \bibnamefont {Kim}}, \bibinfo {author}
  {\bibfnamefont {N.}~\bibnamefont {Lee}}, \ and\ \bibinfo {author}
  {\bibfnamefont {Y.~J.}\ \bibnamefont {Choi}},\ }\href@noop {} {\bibfield
  {journal} {\bibinfo  {journal} {Scientific reports}\ }\textbf {\bibinfo
  {volume} {13}},\ \bibinfo {pages} {7105} (\bibinfo {year}
  {2023})}\BibitemShut {NoStop}%
\bibitem [{\citenamefont {Nickel}\ \emph {et~al.}(2024)\citenamefont {Nickel},
  \citenamefont {Sun}, \citenamefont {Meira}, \citenamefont {Shafer},\ and\
  \citenamefont {van Lierop}}]{nickel2024insights}%
  \BibitemOpen
  \bibfield  {author} {\bibinfo {author} {\bibfnamefont {R.}~\bibnamefont
  {Nickel}}, \bibinfo {author} {\bibfnamefont {C.}~\bibnamefont {Sun}},
  \bibinfo {author} {\bibfnamefont {D.~M.}\ \bibnamefont {Meira}}, \bibinfo
  {author} {\bibfnamefont {P.}~\bibnamefont {Shafer}}, \ and\ \bibinfo {author}
  {\bibfnamefont {J.}~\bibnamefont {van Lierop}},\ }\href@noop {} {\bibfield
  {journal} {\bibinfo  {journal} {Physical Review Materials}\ }\textbf
  {\bibinfo {volume} {8}},\ \bibinfo {pages} {024407} (\bibinfo {year}
  {2024})}\BibitemShut {NoStop}%
\bibitem [{\citenamefont {Gich}\ \emph {et~al.}(2006)\citenamefont {Gich},
  \citenamefont {Frontera}, \citenamefont {Roig}, \citenamefont {Taboada},
  \citenamefont {Molins}, \citenamefont {Rechenberg}, \citenamefont {Ardisson},
  \citenamefont {Macedo}, \citenamefont {Ritter}, \citenamefont {Hardy} \emph
  {et~al.}}]{gich2006high}%
  \BibitemOpen
  \bibfield  {author} {\bibinfo {author} {\bibfnamefont {M.}~\bibnamefont
  {Gich}}, \bibinfo {author} {\bibfnamefont {C.}~\bibnamefont {Frontera}},
  \bibinfo {author} {\bibfnamefont {A.}~\bibnamefont {Roig}}, \bibinfo {author}
  {\bibfnamefont {E.}~\bibnamefont {Taboada}}, \bibinfo {author} {\bibfnamefont
  {E.}~\bibnamefont {Molins}}, \bibinfo {author} {\bibfnamefont
  {H.}~\bibnamefont {Rechenberg}}, \bibinfo {author} {\bibfnamefont
  {J.}~\bibnamefont {Ardisson}}, \bibinfo {author} {\bibfnamefont
  {W.}~\bibnamefont {Macedo}}, \bibinfo {author} {\bibfnamefont
  {C.}~\bibnamefont {Ritter}}, \bibinfo {author} {\bibfnamefont
  {V.}~\bibnamefont {Hardy}},  \emph {et~al.},\ }\href@noop {} {\bibfield
  {journal} {\bibinfo  {journal} {Chemistry of Materials}\ }\textbf {\bibinfo
  {volume} {18}},\ \bibinfo {pages} {3889} (\bibinfo {year}
  {2006})}\BibitemShut {NoStop}%
\bibitem [{\citenamefont {Garc{\'\i}a-Mu{\~n}oz}\ \emph
  {et~al.}(2017)\citenamefont {Garc{\'\i}a-Mu{\~n}oz}, \citenamefont
  {Romaguera}, \citenamefont {Fauth}, \citenamefont {Nogu{\'e}s},\ and\
  \citenamefont {Gich}}]{garcia2017unveiling}%
  \BibitemOpen
  \bibfield  {author} {\bibinfo {author} {\bibfnamefont {J.~L.}\ \bibnamefont
  {Garc{\'\i}a-Mu{\~n}oz}}, \bibinfo {author} {\bibfnamefont {A.}~\bibnamefont
  {Romaguera}}, \bibinfo {author} {\bibfnamefont {F.}~\bibnamefont {Fauth}},
  \bibinfo {author} {\bibfnamefont {J.}~\bibnamefont {Nogu{\'e}s}}, \ and\
  \bibinfo {author} {\bibfnamefont {M.}~\bibnamefont {Gich}},\ }\href@noop {}
  {\bibfield  {journal} {\bibinfo  {journal} {Chemistry of Materials}\ }\textbf
  {\bibinfo {volume} {29}},\ \bibinfo {pages} {9705} (\bibinfo {year}
  {2017})}\BibitemShut {NoStop}%
\bibitem [{\citenamefont {Tseng}\ \emph {et~al.}(2009)\citenamefont {Tseng},
  \citenamefont {Souza-Neto}, \citenamefont {Haskel}, \citenamefont {Gich},
  \citenamefont {Frontera}, \citenamefont {Roig}, \citenamefont
  {Van~Veenendaal},\ and\ \citenamefont {Nogu{\'e}s}}]{tseng2009nonzero}%
  \BibitemOpen
  \bibfield  {author} {\bibinfo {author} {\bibfnamefont {Y.-C.}\ \bibnamefont
  {Tseng}}, \bibinfo {author} {\bibfnamefont {N.~M.}\ \bibnamefont
  {Souza-Neto}}, \bibinfo {author} {\bibfnamefont {D.}~\bibnamefont {Haskel}},
  \bibinfo {author} {\bibfnamefont {M.}~\bibnamefont {Gich}}, \bibinfo {author}
  {\bibfnamefont {C.}~\bibnamefont {Frontera}}, \bibinfo {author}
  {\bibfnamefont {A.}~\bibnamefont {Roig}}, \bibinfo {author} {\bibfnamefont
  {M.}~\bibnamefont {Van~Veenendaal}}, \ and\ \bibinfo {author} {\bibfnamefont
  {J.}~\bibnamefont {Nogu{\'e}s}},\ }\href@noop {} {\bibfield  {journal}
  {\bibinfo  {journal} {Physical Review B}\ }\textbf {\bibinfo {volume} {79}},\
  \bibinfo {pages} {094404} (\bibinfo {year} {2009})}\BibitemShut {NoStop}%
\bibitem [{\citenamefont {Nickel}\ \emph {et~al.}(2023)\citenamefont {Nickel},
  \citenamefont {Gibbs}, \citenamefont {Burgess}, \citenamefont {Shafer},
  \citenamefont {Meira}, \citenamefont {Sun},\ and\ \citenamefont {van
  Lierop}}]{nickel2023nanoscale}%
  \BibitemOpen
  \bibfield  {author} {\bibinfo {author} {\bibfnamefont {R.}~\bibnamefont
  {Nickel}}, \bibinfo {author} {\bibfnamefont {J.}~\bibnamefont {Gibbs}},
  \bibinfo {author} {\bibfnamefont {J.}~\bibnamefont {Burgess}}, \bibinfo
  {author} {\bibfnamefont {P.}~\bibnamefont {Shafer}}, \bibinfo {author}
  {\bibfnamefont {D.~M.}\ \bibnamefont {Meira}}, \bibinfo {author}
  {\bibfnamefont {C.}~\bibnamefont {Sun}}, \ and\ \bibinfo {author}
  {\bibfnamefont {J.}~\bibnamefont {van Lierop}},\ }\href {\doibase
  10.1021/acs.nanolett.3c01512} {\bibfield  {journal} {\bibinfo  {journal}
  {Nano Letters}\ }\textbf {\bibinfo {volume} {23}},\ \bibinfo {pages} {7845}
  (\bibinfo {year} {2023})},\ \bibinfo {note} {pMID: 37625017},\ \Eprint
  {http://arxiv.org/abs/https://doi.org/10.1021/acs.nanolett.3c01512}
  {https://doi.org/10.1021/acs.nanolett.3c01512} \BibitemShut {NoStop}%
\bibitem [{\citenamefont {Tronc}\ \emph {et~al.}(2005)\citenamefont {Tronc},
  \citenamefont {Chan{\'e}ac}, \citenamefont {Jolivet},\ and\ \citenamefont
  {Greneche}}]{tronc2005spin}%
  \BibitemOpen
  \bibfield  {author} {\bibinfo {author} {\bibfnamefont {E.}~\bibnamefont
  {Tronc}}, \bibinfo {author} {\bibfnamefont {C.}~\bibnamefont {Chan{\'e}ac}},
  \bibinfo {author} {\bibfnamefont {J.}~\bibnamefont {Jolivet}}, \ and\
  \bibinfo {author} {\bibfnamefont {J.}~\bibnamefont {Greneche}},\ }\href@noop
  {} {\bibfield  {journal} {\bibinfo  {journal} {Journal of applied physics}\
  }\textbf {\bibinfo {volume} {98}},\ \bibinfo {pages} {053901} (\bibinfo
  {year} {2005})}\BibitemShut {NoStop}%
\bibitem [{\citenamefont {Rehspringer}\ \emph {et~al.}(2006)\citenamefont
  {Rehspringer}, \citenamefont {Vilminot}, \citenamefont {Niznansky},
  \citenamefont {Zaveta}, \citenamefont {Estourn{\`e}s},\ and\ \citenamefont
  {Kurmoo}}]{rehspringer2006temperature}%
  \BibitemOpen
  \bibfield  {author} {\bibinfo {author} {\bibfnamefont {J.-L.}\ \bibnamefont
  {Rehspringer}}, \bibinfo {author} {\bibfnamefont {S.}~\bibnamefont
  {Vilminot}}, \bibinfo {author} {\bibfnamefont {D.}~\bibnamefont {Niznansky}},
  \bibinfo {author} {\bibfnamefont {K.}~\bibnamefont {Zaveta}}, \bibinfo
  {author} {\bibfnamefont {C.}~\bibnamefont {Estourn{\`e}s}}, \ and\ \bibinfo
  {author} {\bibfnamefont {M.}~\bibnamefont {Kurmoo}},\ }in\ \href@noop {}
  {\emph {\bibinfo {booktitle} {ICAME 2005}}}\ (\bibinfo  {publisher}
  {Springer},\ \bibinfo {year} {2006})\ pp.\ \bibinfo {pages}
  {475--481}\BibitemShut {NoStop}%
\bibitem [{\citenamefont {Kohout}\ \emph {et~al.}(2015)\citenamefont {Kohout},
  \citenamefont {Br{\'a}zda}, \citenamefont {Z{\'a}v{\v{e}}ta}, \citenamefont
  {Kub{\'a}niov{\'a}}, \citenamefont {Kmje{\v{c}}}, \citenamefont
  {Kub{\'\i}{\v{c}}kov{\'a}}, \citenamefont {Klementov{\'a}}, \citenamefont
  {{\v{S}}antav{\'a}},\ and\ \citenamefont
  {Lan{\v{c}}ok}}]{kohout2015magnetic}%
  \BibitemOpen
  \bibfield  {author} {\bibinfo {author} {\bibfnamefont {J.}~\bibnamefont
  {Kohout}}, \bibinfo {author} {\bibfnamefont {P.}~\bibnamefont {Br{\'a}zda}},
  \bibinfo {author} {\bibfnamefont {K.}~\bibnamefont {Z{\'a}v{\v{e}}ta}},
  \bibinfo {author} {\bibfnamefont {D.}~\bibnamefont {Kub{\'a}niov{\'a}}},
  \bibinfo {author} {\bibfnamefont {T.}~\bibnamefont {Kmje{\v{c}}}}, \bibinfo
  {author} {\bibfnamefont {L.}~\bibnamefont {Kub{\'\i}{\v{c}}kov{\'a}}},
  \bibinfo {author} {\bibfnamefont {M.}~\bibnamefont {Klementov{\'a}}},
  \bibinfo {author} {\bibfnamefont {E.}~\bibnamefont {{\v{S}}antav{\'a}}}, \
  and\ \bibinfo {author} {\bibfnamefont {A.}~\bibnamefont {Lan{\v{c}}ok}},\
  }\href@noop {} {\bibfield  {journal} {\bibinfo  {journal} {Journal of Applied
  Physics}\ }\textbf {\bibinfo {volume} {117}},\ \bibinfo {pages} {17D505}
  (\bibinfo {year} {2015})}\BibitemShut {NoStop}%
\bibitem [{\citenamefont {Jones}\ \emph {et~al.}(2019)\citenamefont {Jones},
  \citenamefont {Nickel}, \citenamefont {Manna}, \citenamefont {Hilman},\ and\
  \citenamefont {van Lierop}}]{jones2019temperature}%
  \BibitemOpen
  \bibfield  {author} {\bibinfo {author} {\bibfnamefont {R.}~\bibnamefont
  {Jones}}, \bibinfo {author} {\bibfnamefont {R.}~\bibnamefont {Nickel}},
  \bibinfo {author} {\bibfnamefont {P.~K.}\ \bibnamefont {Manna}}, \bibinfo
  {author} {\bibfnamefont {J.}~\bibnamefont {Hilman}}, \ and\ \bibinfo {author}
  {\bibfnamefont {J.}~\bibnamefont {van Lierop}},\ }\href@noop {} {\bibfield
  {journal} {\bibinfo  {journal} {Physical Review B}\ }\textbf {\bibinfo
  {volume} {100}},\ \bibinfo {pages} {094425} (\bibinfo {year}
  {2019})}\BibitemShut {NoStop}%
\bibitem [{\citenamefont {Kimura}\ \emph {et~al.}(2000)\citenamefont {Kimura},
  \citenamefont {Kumai}, \citenamefont {Okimoto}, \citenamefont {Tomioka},\
  and\ \citenamefont {Tokura}}]{kimura2000variation}%
  \BibitemOpen
  \bibfield  {author} {\bibinfo {author} {\bibfnamefont {T.}~\bibnamefont
  {Kimura}}, \bibinfo {author} {\bibfnamefont {R.}~\bibnamefont {Kumai}},
  \bibinfo {author} {\bibfnamefont {Y.}~\bibnamefont {Okimoto}}, \bibinfo
  {author} {\bibfnamefont {Y.}~\bibnamefont {Tomioka}}, \ and\ \bibinfo
  {author} {\bibfnamefont {Y.}~\bibnamefont {Tokura}},\ }\href@noop {}
  {\bibfield  {journal} {\bibinfo  {journal} {Physical Review B}\ }\textbf
  {\bibinfo {volume} {62}},\ \bibinfo {pages} {15021} (\bibinfo {year}
  {2000})}\BibitemShut {NoStop}%
\bibitem [{\citenamefont {Hong}\ \emph {et~al.}(2005)\citenamefont {Hong},
  \citenamefont {Hur},\ and\ \citenamefont {Choi}}]{hong2005magnetic}%
  \BibitemOpen
  \bibfield  {author} {\bibinfo {author} {\bibfnamefont {C.~S.}\ \bibnamefont
  {Hong}}, \bibinfo {author} {\bibfnamefont {N.~H.}\ \bibnamefont {Hur}}, \
  and\ \bibinfo {author} {\bibfnamefont {Y.~N.}\ \bibnamefont {Choi}},\
  }\href@noop {} {\bibfield  {journal} {\bibinfo  {journal} {Solid state
  communications}\ }\textbf {\bibinfo {volume} {133}},\ \bibinfo {pages} {531}
  (\bibinfo {year} {2005})}\BibitemShut {NoStop}%
\bibitem [{\citenamefont {Raies}\ \emph {et~al.}(2020)\citenamefont {Raies},
  \citenamefont {Al~Dulmani}, \citenamefont {Ben~Farhat}, \citenamefont
  {Fadlallah},\ and\ \citenamefont {Amami}}]{raies2020temperature}%
  \BibitemOpen
  \bibfield  {author} {\bibinfo {author} {\bibfnamefont {I.}~\bibnamefont
  {Raies}}, \bibinfo {author} {\bibfnamefont {S.~A.}\ \bibnamefont
  {Al~Dulmani}}, \bibinfo {author} {\bibfnamefont {L.}~\bibnamefont
  {Ben~Farhat}}, \bibinfo {author} {\bibfnamefont {E.~E.}\ \bibnamefont
  {Fadlallah}}, \ and\ \bibinfo {author} {\bibfnamefont {M.}~\bibnamefont
  {Amami}},\ }\href@noop {} {\bibfield  {journal} {\bibinfo  {journal} {Journal
  of Asian Ceramic Societies}\ }\textbf {\bibinfo {volume} {8}},\ \bibinfo
  {pages} {1095} (\bibinfo {year} {2020})}\BibitemShut {NoStop}%
\bibitem [{\citenamefont {Kumar}\ \emph {et~al.}(2021)\citenamefont {Kumar},
  \citenamefont {Warshi}, \citenamefont {Sagdeo},\ and\ \citenamefont
  {Sagdeo}}]{kumar2021investigations}%
  \BibitemOpen
  \bibfield  {author} {\bibinfo {author} {\bibfnamefont {A.}~\bibnamefont
  {Kumar}}, \bibinfo {author} {\bibfnamefont {M.~K.}\ \bibnamefont {Warshi}},
  \bibinfo {author} {\bibfnamefont {A.}~\bibnamefont {Sagdeo}}, \ and\ \bibinfo
  {author} {\bibfnamefont {P.}~\bibnamefont {Sagdeo}},\ }\href@noop {}
  {\bibfield  {journal} {\bibinfo  {journal} {The Journal of Physical Chemistry
  C}\ }\textbf {\bibinfo {volume} {125}},\ \bibinfo {pages} {14048} (\bibinfo
  {year} {2021})}\BibitemShut {NoStop}%
\bibitem [{\citenamefont {Tiburcio}\ \emph {et~al.}(2023)\citenamefont
  {Tiburcio}, \citenamefont {Sacari}, \citenamefont {Chacaltana}, \citenamefont
  {Medina}, \citenamefont {Gamarra}, \citenamefont {Polo}, \citenamefont
  {Mamani},\ and\ \citenamefont {Quispe}}]{tiburcio2023influence}%
  \BibitemOpen
  \bibfield  {author} {\bibinfo {author} {\bibfnamefont {J.}~\bibnamefont
  {Tiburcio}}, \bibinfo {author} {\bibfnamefont {E.}~\bibnamefont {Sacari}},
  \bibinfo {author} {\bibfnamefont {J.}~\bibnamefont {Chacaltana}}, \bibinfo
  {author} {\bibfnamefont {J.}~\bibnamefont {Medina}}, \bibinfo {author}
  {\bibfnamefont {F.}~\bibnamefont {Gamarra}}, \bibinfo {author} {\bibfnamefont
  {C.}~\bibnamefont {Polo}}, \bibinfo {author} {\bibfnamefont {E.}~\bibnamefont
  {Mamani}}, \ and\ \bibinfo {author} {\bibfnamefont {A.}~\bibnamefont
  {Quispe}},\ }\href@noop {} {\bibfield  {journal} {\bibinfo  {journal}
  {Energies}\ }\textbf {\bibinfo {volume} {16}},\ \bibinfo {pages} {786}
  (\bibinfo {year} {2023})}\BibitemShut {NoStop}%
\bibitem [{\citenamefont {Kummer}\ \emph {et~al.}(2016)\citenamefont {Kummer},
  \citenamefont {Fondacaro}, \citenamefont {Jimenez}, \citenamefont
  {Velez-Fort}, \citenamefont {Amorese}, \citenamefont {Aspbury}, \citenamefont
  {Yakhou-Harris}, \citenamefont {Van Der~Linden},\ and\ \citenamefont
  {Brookes}}]{kummer2016high}%
  \BibitemOpen
  \bibfield  {author} {\bibinfo {author} {\bibfnamefont {K.}~\bibnamefont
  {Kummer}}, \bibinfo {author} {\bibfnamefont {A.}~\bibnamefont {Fondacaro}},
  \bibinfo {author} {\bibfnamefont {E.}~\bibnamefont {Jimenez}}, \bibinfo
  {author} {\bibfnamefont {E.}~\bibnamefont {Velez-Fort}}, \bibinfo {author}
  {\bibfnamefont {A.}~\bibnamefont {Amorese}}, \bibinfo {author} {\bibfnamefont
  {M.}~\bibnamefont {Aspbury}}, \bibinfo {author} {\bibfnamefont
  {F.}~\bibnamefont {Yakhou-Harris}}, \bibinfo {author} {\bibfnamefont
  {P.}~\bibnamefont {Van Der~Linden}}, \ and\ \bibinfo {author} {\bibfnamefont
  {N.}~\bibnamefont {Brookes}},\ }\href@noop {} {\bibfield  {journal} {\bibinfo
   {journal} {Synchrotron Radiation}\ }\textbf {\bibinfo {volume} {23}},\
  \bibinfo {pages} {464} (\bibinfo {year} {2016})}\BibitemShut {NoStop}%
\bibitem [{\citenamefont {Brookes}\ \emph {et~al.}(2018)\citenamefont
  {Brookes}, \citenamefont {Yakhou-Harris}, \citenamefont {Kummer},
  \citenamefont {Fondacaro}, \citenamefont {Cezar}, \citenamefont {Betto},
  \citenamefont {Velez-Fort}, \citenamefont {Amorese}, \citenamefont
  {Ghiringhelli}, \citenamefont {Braicovich} \emph
  {et~al.}}]{brookes2018beamline}%
  \BibitemOpen
  \bibfield  {author} {\bibinfo {author} {\bibfnamefont {N.}~\bibnamefont
  {Brookes}}, \bibinfo {author} {\bibfnamefont {F.}~\bibnamefont
  {Yakhou-Harris}}, \bibinfo {author} {\bibfnamefont {K.}~\bibnamefont
  {Kummer}}, \bibinfo {author} {\bibfnamefont {A.}~\bibnamefont {Fondacaro}},
  \bibinfo {author} {\bibfnamefont {J.}~\bibnamefont {Cezar}}, \bibinfo
  {author} {\bibfnamefont {D.}~\bibnamefont {Betto}}, \bibinfo {author}
  {\bibfnamefont {E.}~\bibnamefont {Velez-Fort}}, \bibinfo {author}
  {\bibfnamefont {A.}~\bibnamefont {Amorese}}, \bibinfo {author} {\bibfnamefont
  {G.}~\bibnamefont {Ghiringhelli}}, \bibinfo {author} {\bibfnamefont
  {L.}~\bibnamefont {Braicovich}},  \emph {et~al.},\ }\href@noop {} {\bibfield
  {journal} {\bibinfo  {journal} {Nuclear Instruments and Methods in Physics
  Research Section A: Accelerators, Spectrometers, Detectors and Associated
  Equipment}\ }\textbf {\bibinfo {volume} {903}},\ \bibinfo {pages} {175}
  (\bibinfo {year} {2018})}\BibitemShut {NoStop}%
\bibitem [{\citenamefont {G{\"u}tlich}\ \emph {et~al.}(2010)\citenamefont
  {G{\"u}tlich}, \citenamefont {Bill},\ and\ \citenamefont
  {Trautwein}}]{gutlich2010mossbauer}%
  \BibitemOpen
  \bibfield  {author} {\bibinfo {author} {\bibfnamefont {P.}~\bibnamefont
  {G{\"u}tlich}}, \bibinfo {author} {\bibfnamefont {E.}~\bibnamefont {Bill}}, \
  and\ \bibinfo {author} {\bibfnamefont {A.~X.}\ \bibnamefont {Trautwein}},\
  }\href@noop {} {\emph {\bibinfo {title} {M{\"o}ssbauer spectroscopy and
  transition metal chemistry: {Fundamentals} and applications}}}\ (\bibinfo
  {publisher} {Springer Science \& Business Media},\ \bibinfo {year}
  {2010})\BibitemShut {NoStop}%
\bibitem [{\citenamefont {Potze}\ \emph {et~al.}(1995)\citenamefont {Potze},
  \citenamefont {Sawatzky},\ and\ \citenamefont
  {Abbate}}]{potze1995possibility}%
  \BibitemOpen
  \bibfield  {author} {\bibinfo {author} {\bibfnamefont {R.}~\bibnamefont
  {Potze}}, \bibinfo {author} {\bibfnamefont {G.}~\bibnamefont {Sawatzky}}, \
  and\ \bibinfo {author} {\bibfnamefont {M.}~\bibnamefont {Abbate}},\
  }\href@noop {} {\bibfield  {journal} {\bibinfo  {journal} {Physical Review
  B}\ }\textbf {\bibinfo {volume} {51}},\ \bibinfo {pages} {11501} (\bibinfo
  {year} {1995})}\BibitemShut {NoStop}%
\bibitem [{\citenamefont {Zhuang}\ \emph {et~al.}(1998)\citenamefont {Zhuang},
  \citenamefont {Zhang}, \citenamefont {Hu},\ and\ \citenamefont
  {Ming}}]{zhuang1998possible}%
  \BibitemOpen
  \bibfield  {author} {\bibinfo {author} {\bibfnamefont {M.}~\bibnamefont
  {Zhuang}}, \bibinfo {author} {\bibfnamefont {W.}~\bibnamefont {Zhang}},
  \bibinfo {author} {\bibfnamefont {A.}~\bibnamefont {Hu}}, \ and\ \bibinfo
  {author} {\bibfnamefont {N.}~\bibnamefont {Ming}},\ }\href@noop {} {\bibfield
   {journal} {\bibinfo  {journal} {Physical Review B}\ }\textbf {\bibinfo
  {volume} {57}},\ \bibinfo {pages} {13655} (\bibinfo {year}
  {1998})}\BibitemShut {NoStop}%
\bibitem [{\citenamefont {Pouchard}\ \emph {et~al.}(2001)\citenamefont
  {Pouchard}, \citenamefont {Villesuzanne},\ and\ \citenamefont
  {Doumerc}}]{pouchard2001spin}%
  \BibitemOpen
  \bibfield  {author} {\bibinfo {author} {\bibfnamefont {M.}~\bibnamefont
  {Pouchard}}, \bibinfo {author} {\bibfnamefont {A.}~\bibnamefont
  {Villesuzanne}}, \ and\ \bibinfo {author} {\bibfnamefont {J.-P.}\
  \bibnamefont {Doumerc}},\ }\href@noop {} {\bibfield  {journal} {\bibinfo
  {journal} {Journal of Solid State Chemistry}\ }\textbf {\bibinfo {volume}
  {162}},\ \bibinfo {pages} {282} (\bibinfo {year} {2001})}\BibitemShut
  {NoStop}%
\bibitem [{\citenamefont {Rogge}\ \emph {et~al.}(2018)\citenamefont {Rogge},
  \citenamefont {Chandrasena}, \citenamefont {Cammarata}, \citenamefont
  {Green}, \citenamefont {Shafer}, \citenamefont {Lefler}, \citenamefont
  {Huon}, \citenamefont {Arab}, \citenamefont {Arenholz}, \citenamefont {Lee}
  \emph {et~al.}}]{rogge2018electronic}%
  \BibitemOpen
  \bibfield  {author} {\bibinfo {author} {\bibfnamefont {P.~C.}\ \bibnamefont
  {Rogge}}, \bibinfo {author} {\bibfnamefont {R.~U.}\ \bibnamefont
  {Chandrasena}}, \bibinfo {author} {\bibfnamefont {A.}~\bibnamefont
  {Cammarata}}, \bibinfo {author} {\bibfnamefont {R.~J.}\ \bibnamefont
  {Green}}, \bibinfo {author} {\bibfnamefont {P.}~\bibnamefont {Shafer}},
  \bibinfo {author} {\bibfnamefont {B.~M.}\ \bibnamefont {Lefler}}, \bibinfo
  {author} {\bibfnamefont {A.}~\bibnamefont {Huon}}, \bibinfo {author}
  {\bibfnamefont {A.}~\bibnamefont {Arab}}, \bibinfo {author} {\bibfnamefont
  {E.}~\bibnamefont {Arenholz}}, \bibinfo {author} {\bibfnamefont {H.~N.}\
  \bibnamefont {Lee}},  \emph {et~al.},\ }\href@noop {} {\bibfield  {journal}
  {\bibinfo  {journal} {Physical review materials}\ }\textbf {\bibinfo {volume}
  {2}},\ \bibinfo {pages} {015002} (\bibinfo {year} {2018})}\BibitemShut
  {NoStop}%
\bibitem [{\citenamefont {Korotin}\ \emph {et~al.}(1996)\citenamefont
  {Korotin}, \citenamefont {Ezhov}, \citenamefont {Solovyev}, \citenamefont
  {Anisimov}, \citenamefont {Khomskii},\ and\ \citenamefont
  {Sawatzky}}]{korotin1996intermediate}%
  \BibitemOpen
  \bibfield  {author} {\bibinfo {author} {\bibfnamefont {M.}~\bibnamefont
  {Korotin}}, \bibinfo {author} {\bibfnamefont {S.~Y.}\ \bibnamefont {Ezhov}},
  \bibinfo {author} {\bibfnamefont {I.}~\bibnamefont {Solovyev}}, \bibinfo
  {author} {\bibfnamefont {V.}~\bibnamefont {Anisimov}}, \bibinfo {author}
  {\bibfnamefont {D.}~\bibnamefont {Khomskii}}, \ and\ \bibinfo {author}
  {\bibfnamefont {G.}~\bibnamefont {Sawatzky}},\ }\href@noop {} {\bibfield
  {journal} {\bibinfo  {journal} {Physical Review B}\ }\textbf {\bibinfo
  {volume} {54}},\ \bibinfo {pages} {5309} (\bibinfo {year}
  {1996})}\BibitemShut {NoStop}%
\bibitem [{\citenamefont {Johnston}\ \emph {et~al.}(2014)\citenamefont
  {Johnston}, \citenamefont {Mukherjee}, \citenamefont {Elfimov}, \citenamefont
  {Berciu},\ and\ \citenamefont {Sawatzky}}]{johnston2014charge}%
  \BibitemOpen
  \bibfield  {author} {\bibinfo {author} {\bibfnamefont {S.}~\bibnamefont
  {Johnston}}, \bibinfo {author} {\bibfnamefont {A.}~\bibnamefont {Mukherjee}},
  \bibinfo {author} {\bibfnamefont {I.}~\bibnamefont {Elfimov}}, \bibinfo
  {author} {\bibfnamefont {M.}~\bibnamefont {Berciu}}, \ and\ \bibinfo {author}
  {\bibfnamefont {G.~A.}\ \bibnamefont {Sawatzky}},\ }\href@noop {} {\bibfield
  {journal} {\bibinfo  {journal} {Physical review letters}\ }\textbf {\bibinfo
  {volume} {112}},\ \bibinfo {pages} {106404} (\bibinfo {year}
  {2014})}\BibitemShut {NoStop}%
\bibitem [{\citenamefont {Bisogni}\ \emph {et~al.}(2016)\citenamefont
  {Bisogni}, \citenamefont {Catalano}, \citenamefont {Green}, \citenamefont
  {Gibert}, \citenamefont {Scherwitzl}, \citenamefont {Huang}, \citenamefont
  {Strocov}, \citenamefont {Zubko}, \citenamefont {Balandeh}, \citenamefont
  {Triscone} \emph {et~al.}}]{bisogni2016ground}%
  \BibitemOpen
  \bibfield  {author} {\bibinfo {author} {\bibfnamefont {V.}~\bibnamefont
  {Bisogni}}, \bibinfo {author} {\bibfnamefont {S.}~\bibnamefont {Catalano}},
  \bibinfo {author} {\bibfnamefont {R.~J.}\ \bibnamefont {Green}}, \bibinfo
  {author} {\bibfnamefont {M.}~\bibnamefont {Gibert}}, \bibinfo {author}
  {\bibfnamefont {R.}~\bibnamefont {Scherwitzl}}, \bibinfo {author}
  {\bibfnamefont {Y.}~\bibnamefont {Huang}}, \bibinfo {author} {\bibfnamefont
  {V.~N.}\ \bibnamefont {Strocov}}, \bibinfo {author} {\bibfnamefont
  {P.}~\bibnamefont {Zubko}}, \bibinfo {author} {\bibfnamefont
  {S.}~\bibnamefont {Balandeh}}, \bibinfo {author} {\bibfnamefont {J.-M.}\
  \bibnamefont {Triscone}},  \emph {et~al.},\ }\href@noop {} {\bibfield
  {journal} {\bibinfo  {journal} {Nature Communications}\ }\textbf {\bibinfo
  {volume} {7}},\ \bibinfo {pages} {13017} (\bibinfo {year}
  {2016})}\BibitemShut {NoStop}%
\bibitem [{\citenamefont {Mizokawa}\ \emph {et~al.}(2000)\citenamefont
  {Mizokawa}, \citenamefont {Khomskii},\ and\ \citenamefont
  {Sawatzky}}]{mizokawa2000spin}%
  \BibitemOpen
  \bibfield  {author} {\bibinfo {author} {\bibfnamefont {T.}~\bibnamefont
  {Mizokawa}}, \bibinfo {author} {\bibfnamefont {D.}~\bibnamefont {Khomskii}},
  \ and\ \bibinfo {author} {\bibfnamefont {G.}~\bibnamefont {Sawatzky}},\
  }\href@noop {} {\bibfield  {journal} {\bibinfo  {journal} {Physical Review
  B}\ }\textbf {\bibinfo {volume} {61}},\ \bibinfo {pages} {11263} (\bibinfo
  {year} {2000})}\BibitemShut {NoStop}%
\bibitem [{\citenamefont {Green}\ \emph {et~al.}(2016)\citenamefont {Green},
  \citenamefont {Haverkort},\ and\ \citenamefont {Sawatzky}}]{green2016bond}%
  \BibitemOpen
  \bibfield  {author} {\bibinfo {author} {\bibfnamefont {R.~J.}\ \bibnamefont
  {Green}}, \bibinfo {author} {\bibfnamefont {M.~W.}\ \bibnamefont
  {Haverkort}}, \ and\ \bibinfo {author} {\bibfnamefont {G.~A.}\ \bibnamefont
  {Sawatzky}},\ }\href@noop {} {\bibfield  {journal} {\bibinfo  {journal}
  {Physical Review B}\ }\textbf {\bibinfo {volume} {94}},\ \bibinfo {pages}
  {195127} (\bibinfo {year} {2016})}\BibitemShut {NoStop}%
\bibitem [{\citenamefont {Dalpian}\ \emph {et~al.}(2018)\citenamefont
  {Dalpian}, \citenamefont {Liu}, \citenamefont {Varignon}, \citenamefont
  {Bibes},\ and\ \citenamefont {Zunger}}]{dalpian2018bond}%
  \BibitemOpen
  \bibfield  {author} {\bibinfo {author} {\bibfnamefont {G.~M.}\ \bibnamefont
  {Dalpian}}, \bibinfo {author} {\bibfnamefont {Q.}~\bibnamefont {Liu}},
  \bibinfo {author} {\bibfnamefont {J.}~\bibnamefont {Varignon}}, \bibinfo
  {author} {\bibfnamefont {M.}~\bibnamefont {Bibes}}, \ and\ \bibinfo {author}
  {\bibfnamefont {A.}~\bibnamefont {Zunger}},\ }\href@noop {} {\bibfield
  {journal} {\bibinfo  {journal} {Physical Review B}\ }\textbf {\bibinfo
  {volume} {98}},\ \bibinfo {pages} {075135} (\bibinfo {year}
  {2018})}\BibitemShut {NoStop}%
\bibitem [{\citenamefont {Raebiger}\ \emph {et~al.}(2008)\citenamefont
  {Raebiger}, \citenamefont {Lany},\ and\ \citenamefont
  {Zunger}}]{raebiger2008charge}%
  \BibitemOpen
  \bibfield  {author} {\bibinfo {author} {\bibfnamefont {H.}~\bibnamefont
  {Raebiger}}, \bibinfo {author} {\bibfnamefont {S.}~\bibnamefont {Lany}}, \
  and\ \bibinfo {author} {\bibfnamefont {A.}~\bibnamefont {Zunger}},\
  }\href@noop {} {\bibfield  {journal} {\bibinfo  {journal} {Nature}\ }\textbf
  {\bibinfo {volume} {453}},\ \bibinfo {pages} {763} (\bibinfo {year}
  {2008})}\BibitemShut {NoStop}%
\bibitem [{\citenamefont {Leonov}\ \emph {et~al.}(2006)\citenamefont {Leonov},
  \citenamefont {Yaresko}, \citenamefont {Antonov},\ and\ \citenamefont
  {Anisimov}}]{leonov2006electronic}%
  \BibitemOpen
  \bibfield  {author} {\bibinfo {author} {\bibfnamefont {I.}~\bibnamefont
  {Leonov}}, \bibinfo {author} {\bibfnamefont {A.}~\bibnamefont {Yaresko}},
  \bibinfo {author} {\bibfnamefont {V.}~\bibnamefont {Antonov}}, \ and\
  \bibinfo {author} {\bibfnamefont {V.}~\bibnamefont {Anisimov}},\ }\href@noop
  {} {\bibfield  {journal} {\bibinfo  {journal} {Physical Review B—Condensed
  Matter and Materials Physics}\ }\textbf {\bibinfo {volume} {74}},\ \bibinfo
  {pages} {165117} (\bibinfo {year} {2006})}\BibitemShut {NoStop}%
\end{thebibliography}%

\end{document}